%% file: main.tex
\documentclass[11pt,letterpaper]{article}

\usepackage[british]{babel}

\usepackage[margin=1in]{geometry}
\usepackage{amsfonts} 
\usepackage{xspace}
\usepackage{amsmath,amssymb,amsthm}
\usepackage{graphicx}
\usepackage{xcolor}
\usepackage[colorlinks=true, allcolors=blue]{hyperref}
\usepackage{cleveref}
\usepackage{todonotes}
\usepackage{mathtools}
\usepackage{enumitem}
\usepackage{algorithm}
\usepackage{algorithmic}
\usepackage{multicol}
\usepackage{multirow}
\usepackage{nicefrac}
\usepackage{tikz}
\usetikzlibrary{calc,decorations.pathreplacing,arrows.meta,shapes.geometric}
\usepackage{setspace}
\usepackage{mleftright}

\makeatletter
\renewcommand{\paragraph}{%
  \@startsection{paragraph}{4}%
  {\z@}{1ex \@plus 1ex \@minus .2ex}{-1em}%
  {\normalfont\normalsize\bfseries}%
}
\renewcommand{\subparagraph}{%
	\@startsection{subparagraph}{5}%
	{\parindent}{1ex \@plus 1ex \@minus .2ex}{-1em}%
	{\normalfont\normalsize\bfseries}%
}
\makeatother

\newtheorem{theorem}{Theorem}[section]
\makeatletter

\newtheorem{lemma}[theorem]{Lemma}

\theoremstyle{definition}
\newtheorem{definition}[theorem]{Definition}

\newcommand{\eps}{\epsilon}

\newcommand{\eat}[1]{}

\newcommand{\R}{\mathbb{R}}
\newcommand{\bS}{\mathbb{S}}
\DeclareMathOperator*{\E}{\ensuremath{\mathbb{E}}}
\newcommand{\Z}{\mathbb{Z}}

\newcommand{\calC}{\mathcal{C}}
\newcommand{\calD}{\mathcal{D}}
\newcommand{\calE}{\mathcal{E}}

\newcommand{\calG}{\mathcal{G}}

\newcommand{\calI}{\mathcal{I}}
\newcommand{\calL}{\mathcal{L}}

\newcommand{\calN}{\mathcal{N}}
\newcommand{\calR}{\mathcal{R}}

\eat{

}

\makeatletter
\newcommand*{\rom}[1]{\expandafter\@slowromancap\romannumeral #1@}
\makeatother

\DeclareMathOperator{\dist}{dist}
\DeclareMathOperator{\supp}{supp}
\DeclareMathOperator{\sgn}{sgn}

\DeclareMathOperator{\Var}{Var}

\DeclarePairedDelimiter{\norm}{\lVert}{\rVert}

\DeclarePairedDelimiterX{\inner}[2]{\langle}{\rangle}{#1 , #2}
\DeclarePairedDelimiterXPP\normp[1]{}{\lVert}{\rVert}{_p}{#1}

\NewDocumentCommand\cost{o}{%
    \IfNoValueTF{#1}%
    {\operatorname{cost}}
    {\operatorname{cost}^{(#1)}}
}

\NewDocumentCommand\Ballh{o}{%
    \IfNoValueTF{#1}%
    {\operatorname{Ball}}
    {\operatorname{Ball}_{H}^{(#1)}}
}

\NewDocumentCommand\Ballsh{o}{%
    \IfNoValueTF{#1}%
    {\operatorname{Balls}}
    {\operatorname{Balls}_{H}^{(#1)}}
}

\allowdisplaybreaks

\title{Near-Tight Robust Coresets for Clustering in Euclidean Spaces}

\author{Pengsen Huang\\  Shanghai Jiaotong University \\ \texttt{2076507920@sjtu.edu.cn}
\and
Yi Li\\ School of Physical and Mathematical Sciences\\ Nanyang Technological University \\ \texttt{yili@ntu.edu.sg}
\and
Xuan Wu\\ John Hopcroft Center for Computer Science\\ Shanghai Jiaotong University \\ \texttt{wuxuan2026@sjtu.edu.cn}}
\date{}

\begin{document}

\pagenumbering{roman} 

\maketitle

\begin{abstract}
This paper considers coresets for the robust $(k,z)$-clustering problem with $m$ outliers in Euclidean spaces. A robust coreset is a weighted subset of the data that, for every set of $k$ centers and every outlier budget $0\leq b\leq m$, approximates the clustering cost with $b$ outliers removed to within relative error $\epsilon$. For every constant $z\geq 1$, we prove that the optimal worst-case size of such coresets over all ambient dimensions is
\begin{align*}
	&\Theta(m) + \widetilde{\Theta}_z(\min\{k\eps^{-(z+2)}, k^{2(z+1)/(z+2)}\eps^{-2} + km^{z/(z+2)}\eps^{-4/(z+2)}\}), & 1\leq z \leq 2, \\
	&\Theta(m) + \widetilde{\Theta}_z(\min\{k\eps^{-(z+2)}, k^{2(z+1)/(z+2)}\eps^{-2} + k \min\{m, \sqrt{m}\eps^{-z/2}\} \}), & z > 2.
\end{align*}
\end{abstract}

\pagenumbering{arabic} 

\input{intro}

\input{ring}
\input{sampling}

\input{lb}

\section*{AI Disclosure}
ChatGPT 6 Pro was used to help develop and refine parts of the proofs, check calculations and revise the exposition. All AI-assisted arguments were independently checked by the authors, who take full responsibility for all mathematical claims and the final manuscript. Claude Opus 5.5 was used to revise the exposition and generate Figure~\ref{fig:hard-instance}.

\bibliographystyle{alphaurl}
\bibliography{ref}

\appendix
\input{vanilla}

\end{document}

%% file: intro.tex
\section{Introduction}

We give near-tight bounds on the size of robust coresets for Euclidean clustering. Given a dataset $X\subset \mathbb{R}^d$, the aim of $(k,z)$-clustering with $m$ outliers is to find a set $\calC$ of $k$ centers minimizing
$$
\cost^{(m)}(X,\calC):=\min_{Y\subseteq X, |X\setminus Y|=m}\cost(Y,\calC),
$$ where $\cost(Y,\calC)=\sum_{x\in Y}\min_{c\in \calC} \|x-c\|_2^z$.

\emph{Coresets} are a popular tool for center-based clustering. For robust clustering, a coreset is a small weighted subset that approximates $\cost[b](X,\calC)$ for every real $0\leq b\leq m$ and every set $\calC$ of $k$ centers (see Section~\ref{sec:preliminaries}).

A series of works~\cite{Cohen2022Towards,cohen2022improved,huang2024optimal,HHLX26} established that the optimal worst-case coreset size for Euclidean $(k,z)$-clustering without outliers is $\tilde{\Theta}(Q_z(k,\eps))$, where $Q_z(k,\eps)=\min\{k^{2(z+1)/(z+2)}\eps^{-2},k\eps^{-z-2}\}$.

The first coreset construction for robust clustering, due to Feldman and Schulman~\cite{feldman2012data}, had size exponential in $k+m$. Huang et al.~\cite{huang2022near} gave a near-linear-time construction of size $O(m)+\mathrm{poly}(k/\eps)$ for fixed $z$ and proved a lower bound of $\Omega(m)$. Subsequent work obtained the vanilla coreset size up to polylogarithmic factors at the cost of a larger outlier term. Up to polylogarithmic factors in $km/\eps$ for $m\geq 1$, this term is $km\eps^{-1}$ in~\cite{pmlr-v267-in25a} and $\min\{km\eps^{-1},m\eps^{-2z}\}$ in~\cite{jiang2025coresets}. Huang et al.~\cite[Theorem~E.1]{HJLW26} obtained a Euclidean bound of $O(m\eps^{-z})+\tilde{O}(Q_z(k,\eps))$ for every fixed $z\geq 1$. As they observe for $k$-median~\cite[Theorem~A.1]{HJLW26}, the guarantee at $b=0$ implies the vanilla lower bound. The same observation for general $z$, together with the $\Omega(m)$ lower bound, gives $\tilde{\Omega}(m+Q_z(k,\eps))$. Moreover, in metric spaces of VC or doubling dimension $d$, they obtain $O(m)+\tilde{O}(kd\eps^{-2z})$, with optimal dependence on $m$~\cite[Theorem~D.1]{HJLW26}. This suggests the conjecture that in Euclidean space the optimal robust coreset size is $\tilde{O}(m+Q_z(k,\eps))$.

In this paper, we disprove this conjecture and determine the optimal robust coreset size up to polylogarithmic factors. Along the way, we also give a near-optimal excess risk bound for robust clustering.

\subsection{Preliminaries}\label{sec:preliminaries}
Throughout this paper, we fix $z\geq 1$. Unless stated to be absolute, constants may depend only on $z$. In our coreset size bounds, $\tilde{O}(\cdot)$ suppresses factors polynomial in $\log(k+1)$ and $\log(1/\eps)$. For two points $x,y\in \mathbb{R}^d$, let $\dist(x,y)=\|x-y\|_2$. For a point $x\in \mathbb{R}^d$ and a finite set $\calC\subset\mathbb{R}^d$, $\dist(x,\calC)=\min_{c\in \calC} \dist(x,c)$. Denote by $B(a,r)$ the closed ball of radius $r$ centered at $a\in \mathbb{R}^d$. For a weighted set $T$, we write $w_T(x)$ for the nonnegative weight of $x$ in $T$, with $w_T(x)=0$ for $x\notin T$. For $A\subseteq T$, we write $w_T(A)=\sum_{x\in A}w_T(x)$. We abbreviate $w_T(x)$ to $w_x$ when the dataset is clear.

In $(k,z)$-clustering, the cost is
\[
	\cost(P,\calC)=\sum_{x\in P} w_P(x)\dist^z(x,\calC),
\]
where $w_P(x)=1$ unless $P$ is specified to be weighted. We suppress the fixed exponent $z$ in the cost notation.

\paragraph{Robust clustering for weighted instances.}
For a weighted dataset $X$ and any real budget $0\leq m\leq w_X(X)$, we allow fractional removal of weight and define

$$
\cost^{(m)}(X,\calC):=\min_{\substack{0\leq a_x\leq w_X(x)\\\sum_{x\in X}a_x=m}}\sum_{x\in X}(w_X(x)-a_x)\dist^z(x,\calC).
$$

For unweighted $X$ and integer $m$, this agrees with the definition in the introduction, since an optimal removal deletes the $m$ farthest points from $\calC$.

\paragraph{Robust coresets.} Let $X$ be an unweighted dataset and let $0\leq m\leq |X|$ be an integer. A weighted subset $S\subseteq X$ is a robust $(\eps,m)$-coreset for $(k,z)$-clustering if $w_S(S)\geq m$ and, for every set $\calC$ of $k$ centers and every real $0\leq b\leq m$,
\[
    |\cost[b](S,\calC)-\cost[b](X,\calC)|\leq \eps\cost[b](X,\calC).
\]

When $m=0$, we call $S$ an $\eps$-coreset for vanilla $(k,z)$-clustering. The guarantee is $|\cost(S,\calC)-\cost(X,\calC)|\leq\eps\cost(X,\calC)$ for every set $\calC$ of $k$ centers.

\paragraph{$(r,k)$-instances and regular instances.} Following \cite{HJLW26}, a dataset $Y$ is an $(r,k)$-instance if it can be partitioned into $g\leq k$ nonempty parts $Y_1,\dots,Y_g$ such that $Y_i\subseteq B(a_i,r)$ for some $a_i\in\mathbb{R}^d$ for every $i\in[g]$. We call $Y$ an $(r,k)$-regular instance if $|Y_i|/|Y|\in [\frac{1}{2g},\frac{2}{g}]$ for every $i\in[g]$.

For a subset $I\subseteq [g]$, we let $Y_I:=\bigcup_{i\in I} Y_i$, and for a subset $Z\subseteq Y$, we let $Z_I:=Z\cap Y_I$.

\paragraph{ReLU on distance.} Let $t_+=\max\{t,0\}$. Let $h_u(x)=(u-x)_+$ and define
\[
    H_{u,\lambda}(X,\calC)=\sum_{x\in X}w_X(x)h_u(\lambda\dist^z(x,\calC)).
\]

The following identity relates the robust cost to the ReLU sums by linear programming duality.
\begin{lemma} \label{lem:dual}
	For $\lambda>0$ and $0\leq m\leq w_X(X)$, $\lambda\cost^{(m)}(X,\calC)=\max_{u\geq 0} \big\{(w_X(X)-m)\cdot u-H_{u,\lambda}(X,\calC)\big\}$.
\end{lemma}

\begin{proof}
	Let $Y$ be the weighted set of inliers obtained by removing the $m$ units of weight farthest from $\calC$. This removal minimizes the remaining cost, so $w_Y(Y)=w_X(X)-m$ and $\cost^{(m)}(X,\calC)=\cost(Y,\calC)$. Hence, for every $u\geq 0$,
	\begin{align*}
		&\quad\ (w_X(X)-m)\cdot u-H_{u,\lambda}(X,\calC)\\
		&=w_Y(Y)u-\sum_{x\in X} w_X(x)(u-\lambda\dist^z(x,\calC))_+\\
		&=\sum_{x\in X} \big(w_Y(x)u-w_X(x)(u-\lambda\dist^z(x,\calC))_+\big)\\
		&= \sum_{x\in X} w_Y(x)\big(u-(u-\lambda\dist^z(x,\calC))_+\big)-\sum_{x\in X}(w_X(x)-w_Y(x))h_u(\lambda\dist^z(x,\calC))\\
		&\leq
		\lambda\sum_{x\in X} w_Y(x)\dist^z(x,\calC)-\sum_{x\in X}(w_X(x)-w_Y(x))h_u(\lambda\dist^z(x,\calC))\\
		&= \lambda\cost^{(m)}(X,\calC)-\sum_{x\in X}(w_X(x)-w_Y(x))h_u(\lambda\dist^z(x,\calC))\\
		&\leq 
		\lambda\cost^{(m)}(X,\calC).
	\end{align*}
	On the other hand, let $u^*=\max\{\lambda\dist^z(x,\calC)\mid w_Y(x)>0\}$, with $u^*=0$ if all weight is removed. Then, when $\lambda\dist^z(x,\calC)>u^*$, we have $w_Y(x)=0$, and when $\lambda\dist^z(x,\calC)<u^*$, we have $w_Y(x)=w_X(x)$. Thus both inequalities above become equalities at $u=u^*$, which completes the proof.
\end{proof}

\paragraph{Excess risk.} Let $\calD$ be a distribution on $B(0,1)$ and $X = (x_1,\dots,x_n)\sim\calD^n$ be a sample of size $n$.
For a set $\calC\subset\mathbb{R}^d$ of $k$ centers, define the quantile function of the loss by
\[
q_{\calC}(v)=\sup \mleft\{t \geq 0 \,\bigg\vert\, \Pr_{x\sim \calD} \{ \dist^z(x,\calC)\geq t \} \geq 1-v \mright\},\qquad 0 < v < 1.
\]
For every $\nu\in [0,1]$, the empirical and population risks are
\[
\hat{\calR}^{(\nu)}(X,\calC):=\frac{1}{n} \cost^{(\nu n)}(X,\calC),
\qquad
\calR^{(\nu)}(\calD,\calC):=\int_{0}^{1-\nu} q_{\calC}(v)\,dv.
\]
Choose a minimizer $\calC_X^*\subseteq B(0,1)$ of $\hat{\calR}^{(\nu)}(X,\calC)$ over sets of $k$ centers. Such a minimizer exists since projecting centers onto $B(0,1)$ can only decrease distances to points in the ball. The expected excess risk is
$$
\calE_n^{(\nu)}=\E_{X\sim\calD^n}\big[\calR^{(\nu)}(\calD,\calC^*_X)\big]-\inf_{\calC\subset\mathbb{R}^d,|\calC|=k} \calR^{(\nu)}(\calD,\calC).
$$

\input{results}

\subsection{Technical Overview}

\input{ub_overview}

\input{lb_overview}

\input{related}

%% file: results.tex
\subsection{Our Results}
Our main theorem is the following.
\begin{theorem} \label{thm:main}
There is a polynomial-time algorithm that, given an unweighted dataset $X\subset \mathbb{R}^d$, a positive integer $k$, an integer outlier budget $0\leq m\leq |X|$ and an accuracy parameter $0<\eps<1/2$, constructs, with probability at least $0.9$, a robust $(\eps,m)$-coreset of size $O(m)+\tilde{O}(Q_z(k,\eps)+F_z(k,m,\eps))$ for $(k,z)$-clustering, where
$Q_z(k,\eps)$ has the equivalent form $k\eps^{-2}\min\{k^{\frac{z}{z+2}},\eps^{-z}\}$ and
	\[
		F_z(k,m,\eps) = 
		\begin{cases}
			\min\{km^{\frac{z}{z+2}}\eps^{-\frac{4}{z+2}},k\eps^{-z-2}\}, & 1\leq z\leq 2, \\
			\min\{km,k\sqrt{m}\eps^{-z/2},k\eps^{-z-2}\}, & z>2.
		\end{cases}
	\]
\end{theorem}

Up to logarithmic factors, $Q_z(k,\eps)$ is the coreset size for vanilla $(k,z)$-clustering (the case $m=0$). Allowing outliers contributes the terms $O(m)$ and $\tilde O(F_z(k,m,\eps))$ to the bound.

We prove the following near-tight lower bound.

\begin{theorem}[Informal statement of Theorems~\ref{thm:global_LB} and~\ref{thm:global_LB_z>2}]
	Let $z\geq 1$ be a constant. For all integers $k\geq 1$ and $m\geq 0$ and all $0<\eps\leq\eps_0(z)$, there exists a finite point set $X$ in a Euclidean space with $|X|>m$ such that every robust $(\eps,m)$-coreset for $(k,z)$-clustering of $X$ has size
	\[
		\Omega(m)+\tilde{\Omega}\bigl(Q_z(k,\eps)+F_z(k,m,\eps)\bigr),
	\]
where $\eps_0(z)>0$ depends only on $z$.
\end{theorem}

We also prove an excess risk bound for robust clustering that matches, up to logarithmic factors, the known bound for vanilla $(k,z)$-clustering~\cite[Theorem~5.1]{BLST23}. For $k$-means, a sharper logarithmic dependence is given in \cite{Cohen25}.

\begin{theorem} \label{thm:geb} 
Let $z\geq 1$. For every distribution $\calD$ on $B(0,1)\subset\mathbb R^d$, every $\nu\in [0,1]$ and all $k,n\geq 1$, it holds that
\[
	\calE_n^{(\nu)} \leq C_z\sqrt\frac{k}{n}\log^{\frac{3}{2}}(2kn),
\]
where $C_z$ is a constant depending only on $z$.
\end{theorem}

%% file: ub_overview.tex
\paragraph{Upper bound.} We follow the framework of \cite{HJLW26} to decompose the dataset into three parts: an outer group, a main group, and an inner group. We handle the outer group and the inner group in the same way as in \cite{HJLW26}, which contributes  $O(m)+\tilde{O}(Q_z(k,\eps))$ points. The key improvement is on the main group.

As in \cite{HJLW26}, we reduce the main group to $\tilde{O}(1)$ many $(r,k)$-regular instances. The existing analysis of uniform sampling over a regular instance in~\cite{HJLW26} gives a suboptimal size bound. Our construction still samples uniformly within each remaining cluster, but our treatment of a regular instance incorporates the following three new ingredients.

(i) We introduce a threshold $t$, and add any cluster containing at most $2m/t$ points directly into the coreset, which adds at most $\tilde{O}(km/t)$ points. A key observation is that over the remaining clusters, fewer than $t$ clusters can lose half their weight under the removal of at most $m$ points. This inspires us to develop a more sensitive concentration inequality for these clusters and balance it against the previous $km/t$ term.

(ii) Our analysis is based on the duality in Lemma~\ref{lem:dual}, which expresses the robust cost through ReLU sums $H_{u,\lambda}$, i.e.\ sums of the functions $(u-\lambda\dist^z(x,\calC))_+$.
Prabhu et al.~\cite{prabhu2027coresets} use a related dual formulation for capacitated clustering.
Fix $\calC$, the number $b$ of outliers and an optimal removal. 
A cluster is \emph{good} if it retains at least half of its weight and \emph{bad} otherwise. On good clusters, the vanilla cost is bounded by the robust cost up to a constant factor and an additive error, so a two-sided error bound on the ReLU sums suffices (Lemma~\ref{lemma:uniform1}). We prove this bound by adapting the chaining argument of~\cite{HJLW26}, and it needs $\tilde{O}(Q_z(k,\eps))$ samples, as in the vanilla case.

(iii) The new concentration inequality from (i) is needed only for the bad clusters, and its proof is the most central and involved part of our upper bound.

For the fewer than $t$ bad clusters, duality only requires an upper bound on $H_{u,1+\eps}(E',\calC)-H_{u,1}(E,\calC)$, where $E$ is whichever of the instance and the sample $S$ has the larger robust cost and $E'$ is the other.
Lemma~\ref{lemma:uniform2} gives this bound with $\tilde{O}(k\eps^{-2}B_z(t,\eps))$ samples ($B_z$ is defined in Section~\ref{sec:upper_bounds}).
The dependence on $t$ comes from the chaining nets, which only need to cover the at most $t$ bad clusters.
The factor $1+\eps$ is what makes the bound possible.
It suffices to consider active bad clusters, i.e., those containing a point with $(1+\eps)\dist^z(x,\calC)<u$; the other bad clusters can be ignored.
For large thresholds, every point $x$ of these clusters satisfies $\dist^z(x,\calC)\asymp_z u$. Writing $a=\dist^z(x,\calC)$, we see that the factor $1+\eps$ lowers the term $(u-a)_+$ by an amount of order $\eps u$ whenever $(u-a)_+\gtrsim \eps u$.
After a delicate discretization and symmetrization, this becomes a subtraction term in the upper bound
\begin{equation}\label{eqn:intro_key_ineq}
	\E \sup_{\calC, u} \left(H_{u,1+\eps}(E',\calC)-H_{u,1}(E,\calC)\right) \lesssim \E \sup_{\calC, u} \left(Z_{\calC, u} - c_z\eps u\, \mu(A_{\calC,u}) \right),
\end{equation}
where both ReLU sums are restricted to the active bad clusters and both suprema range over $(\calC,u)$ with $u$ in a fixed range of comparable large thresholds.
Here $Z_{\calC,u}$ is a Rademacher sum built from a discretized, truncated version $f_{\calC,u}$ of the ReLU function $x\mapsto (u - \dist^z(x,\calC))_+$. Let $A_{\calC,u}$ be its support on $S\cup S'$, where $S'$ is an independent copy of $S$. Then $u-\dist^z(x,\calC)\gtrsim_z\eps u$ for all $x\in A_{\calC,u}$. The quantity $\mu(A_{\calC,u})$ is the total weight of the observations in $A_{\calC,u}$ from both $S$ and $S'$.

Let $n$ be the size of the regular instance and $s$ the sample size. Conditional on $S$ and $S'$, the standard deviation of $Z_{\calC,u}$ for a fixed query $(\calC,u)$ is at most of order $u\sqrt{n\,\mu(A_{\calC,u})/s}$, while the subtraction term, being proportional to $\mu(A_{\calC,u})$, does not depend on $s$.
In the chaining argument for the supremum over all queries, we use finite nets of queries at different scales. However, uniform closeness says nothing about supports, as raising the threshold $u$ slightly changes the function by little but can make it positive on many more points. Thus a query with small $\mu(A_{\calC,u})$ may be approximated by net points with $\mu(A)$ as large as about $n$, and the resulting bound on $Z_{\mathcal{C},u}$ is of order $un\sqrt{L/s}$, where $L$ reflects the complexity of the queries. This decays only like $1/\sqrt{s}$ and is not offset by the small subtraction term of that query.
A localization argument in the spirit of~\cite{Peter02} avoids this by grouping queries $(\calC, u)$ according to $\mu(A_{\calC,u})$ and, for the group with $\mu(A_{\mathcal{C},u})\asymp P$, using nets consisting only of queries with $\mu(A)\lesssim P$. On this group, the supremum of $Z_{\mathcal{C},u}$ is at most of order $u\sqrt{nPL/s}$, while every subtraction term is of order $\epsilon uP$, so the group contributes at most
\[
	\max_{P>0} \left(u\sqrt{nPL/s}-\epsilon uP\right) \asymp \frac{unL}{\epsilon s}.
\]
Summing over the $O(\log s)$ groups, the right-hand side of \eqref{eqn:intro_key_ineq} decays like $1/s$ instead of $1/\sqrt{s}$, allowing for a smaller $s$.

Balancing $\tilde{O}(k\eps^{-2}B_z(t,\eps))$ against the $\tilde O(km/t)$ points added in~(i) gives the term $\tilde{O}(F_z(k,m,\eps))$.

%% file: lb_overview.tex
\paragraph{Lower bound.} The idea is to build an instance made of many cells of points, so that a coreset with too few points leaves many cells insufficiently sampled. Each cell is a copy of a set $\Gamma$ of $M$ unit vectors from Lemma~\ref{lem:Y} such that, for a suitable finite set $V$ of
directions, no weighted subset of $N\asymp\tau^{-2}/\log^3(1/\tau)$ points of $\Gamma$ approximates the sums $\sum_{y\in \Gamma}\phi_{u_0}(\langle y,v\rangle)$ well: on average over $v\in V$, the error is at least $M\tau$.
Here $\phi_{u_0}$ is a perturbed ReLU. The proof follows the Boolean Fourier approach of~\cite{LWW19}. On the hypercube of dimension $q\asymp\log N$, the (symmetrized) kernel matrix of $\phi_{u_0}$ is an $H\times H$ matrix with $H=2^{q-1}$. This matrix is diagonalized by the Walsh characters, and its smallest singular value is $\Omega(\sqrt H/q^{3/2})$. Combined with a count of sign patterns from an extension of Warren's theorem~\cite{GJ95}, this gives a probability measure that is far from every measure on $N$ points, and sampling from it yields $\Gamma$. The factor $q^{3/2}$ causes the $\log^3(1/\tau)$ loss.

\input{hard_instance_figure}

The overall hard instance can be viewed as a grid of $K$ rows and $D$ columns, where each cell contains a copy of $\Gamma$ (Figure~\ref{fig:hard-instance}). As
in~\cite{Cohen2022Towards,huang2024optimal,HHLX26}, each cell is shifted by a direction for its row and a direction for its column and these row and column directions are orthogonal to each other and to the space containing $\Gamma$. 
	
We compare two sets of centers. Under $\calC_0$, a chosen set $J$ of $j$ rows is far from all centers (at distance $u_0$) and each other row is close to its center. The other center set $\calC_1$ adds an extra center $c_\ell$ for each column $\ell$ in a subset $\calL\subseteq [D]$. Write $x_{i\ell y}$ for
the copy of $y\in \Gamma$ in cell $(i,\ell)$. For any subsets $I_\ell\subseteq J$ and unit vectors $v_\ell$, the extra
centers can be placed so that for every point,
\[
\dist^z(x_{i\ell y},\calC_0)-\dist^z(x_{i\ell y},\calC_1)
=\begin{cases}
	z u_0^{z-1}\phi_{u_0}(\langle y,v_\ell\rangle), & \ell\in \calL,\ i\in I_\ell,\\
	0, & \text{otherwise}.
\end{cases}
\]
So the change in cost from $\calC_0$ to $\calC_1$ is a sum of terms $z u_0^{z-1}\sum_{y\in \Gamma}\phi_{u_0}(\langle y,v_\ell\rangle)$, one for each cell $(i,\ell)$ with $\ell\in \calL$ and $i\in I_\ell$ (we call these the \emph{selected cells}). If a coreset has too few points, many cells contain at most $N$ of them, and on these cells it misestimates these sums for a suitable choice of $v_\ell$, by the hardness property of $\Gamma$ described in the preceding paragraph.
We choose $I_\ell$ to keep only the cells where the coreset errs in the same way (all overestimates or all underestimates), so the errors do not cancel and add up to more than $\eps$ times the cost.
So the coreset cannot approximate the costs of both $\calC_0$ and $\calC_1$, and every coreset must have $\Omega(KDN)$ points. Thus, the larger the grid, the stronger the lower bound.

Without outliers, this argument gives the vanilla lower bound (Theorem~\ref{thm:vanilla_LB}), with only a $\log^3$ loss when $k\le\eps^{-(z+2)}$, improving on the $\log^5$ loss of~\cite{HHLX26}. However, each far cell in a column of $\calL$ contributes an error of about $\eps$ times the cell's own cost, while far cells in other columns add cost but no error. For the total error to exceed $\eps$ times the total cost, $\calL$ must therefore contain a constant fraction of the columns. Since each column in $\calL$ and each close row needs its own center, the grid can have at most about $k$ rows and $k$ columns.

With outliers, we select only $|\calL| = K$ of the $D$ columns and can take $D\gg K$. 
The far points outside the selected cells all have the largest cost under both $\calC_0$ and $\calC_1$, so removing outliers among them lowers both costs by the same amount, and they no longer affect the comparison.
The number of columns is then limited by the outlier budget $m$ instead of by $k$. Since $J$ and $\calL$ are chosen after the coreset is fixed, the coreset must still have $\Omega(KDN)$ points in total (Lemma~\ref{lem:construction_support}). Choosing the number of far rows, the grid size and the cell size subject to these limits gives a lower bound of $\tilde{\Omega}(Q_z(k,\eps) + F_z(k,m,\eps))$, which, together with the elementary $\Omega(m)$ bound, matches the upper bound up to logarithmic factors.

%% file: hard_instance_figure.tex
\begin{figure}[t]
	\centering
	\begin{tikzpicture}[
		x=1cm,y=1cm,
		cell/.style={draw=black!45,line width=0.3pt,rounded corners=1pt},
		near/.style={cell,fill=blue!7},
		far/.style={cell,fill=black!12},
		hit/.style={cell,fill=orange!55,draw=orange!80!black},
		pt/.style={circle,fill=black!70,inner sep=0pt,minimum size=1.6pt},
		ctr/.style={circle,fill=blue!70!black,inner sep=0pt,minimum size=4pt},
		cl/.style={diamond,fill=orange!85!black,inner sep=0pt,minimum size=6pt},
		lbl/.style={font=\footnotesize},
		brace/.style={decorate,decoration={brace,amplitude=4pt},line width=0.5pt}
		]
		\def\cw{0.62}\def\ch{0.42}\def\dx{0.72}\def\dy{0.52}
		\def\nrow{7}\def\ncol{12}
		\foreach \i in {1,...,\nrow}{
			\foreach \l in {1,...,\ncol}{
				\pgfmathsetmacro{\x}{(\l-1)*\dx}
				\pgfmathsetmacro{\y}{-(\i-1)*\dy}
				\ifnum\i>4 \def\st{far}\else\def\st{near}\fi
				\ifnum\l=3 \ifnum\i=5 \def\st{hit}\fi \ifnum\i=7 \def\st{hit}\fi \fi
				\ifnum\l=7 \ifnum\i=6 \def\st{hit}\fi \fi
				\ifnum\l=10 \ifnum\i=5 \def\st{hit}\fi \ifnum\i=6 \def\st{hit}\fi \fi
				\draw[\st] (\x,\y) rectangle ++(\cw,\ch);
				\node[pt] at (\x+0.16,\y+0.27) {};
				\node[pt] at (\x+0.31,\y+0.13) {};
				\node[pt] at (\x+0.46,\y+0.29) {};
			}
		}
		\foreach \i in {1,...,4}{
			\pgfmathsetmacro{\y}{-(\i-1)*\dy+0.5*\ch}
			\node[ctr] (p\i) at (-0.35,\y) {};
			\node[lbl,anchor=east] at (-0.42,\y) {$p_{\i}$};
		}
		\pgfmathsetmacro{\ymid}{-5*\dy+0.5*\ch}
		\node[ctr,fill=black!75] (cs) at (-0.6,\ymid) {};
		\node[lbl,anchor=east] at (-0.67,\ymid) {$c_\ast$};
		\foreach \i in {5,6,7}{
			\pgfmathsetmacro{\y}{-(\i-1)*\dy+0.5*\ch}
			\draw[black!60,line width=0.4pt] (cs) -- (-0.05,\y);
		}
		\pgfmathsetmacro{\xr}{(\ncol-1)*\dx+\cw+0.12}
		\draw[brace] (\xr,\ch) -- (\xr,-3*\dy)
		node[midway,right=5pt,lbl,align=left]{$[K]\setminus J$: served by $p_i$\\at distance $1$};
		\draw[brace] (\xr,-4*\dy+\ch) -- (\xr,-6*\dy)
		node[midway,right=5pt,lbl,align=left]{$j$ far rows:\\served by $c_\ast$ at\\distance $u_0$};
		\foreach \l/\nm in {3/1,7/2,10/3}{
			\pgfmathsetmacro{\x}{(\l-1)*\dx+0.5*\cw}
			\node[cl] (c\l) at (\x,-6*\dy-0.42) {};
			\node[lbl,anchor=north] at (\x,-6*\dy-0.5) {$c_{\ell_{\nm}}$};
			\draw[-{Stealth[length=3pt]},orange!80!black,line width=0.5pt]
			(c\l) -- (\x,-6*\dy-0.04);
		}
		\node[lbl,anchor=north west,orange!60!black] at ({(\ncol-1)*\dx+\cw+0.12},-6*\dy-0.3) {extra centers for $\ell\in\calL$};
		\draw[brace] (0,\ch+0.1) -- ({(\ncol-1)*\dx+\cw},\ch+0.1)
		node[midway,above=5pt,lbl]{$D$ columns $\ell$; each cell $X_{i\ell}$ is a copy of $\Gamma$ ($M$ points)};
		\begin{scope}[shift={(0.2,-6*\dy-1.35)}]
			\draw[near] (0,0) rectangle ++(0.32,0.22); \node[lbl,anchor=west] at (0.38,0.11) {near cells (cost $1$ per point)};
			\draw[far] (4.9,0) rectangle ++(0.32,0.22); \node[lbl,anchor=west] at (5.28,0.11) {far cells (mostly removed as outliers)};
			\draw[hit] (0,-0.38) rectangle ++(0.32,0.22); \node[lbl,anchor=west] at (0.38,-0.27) {cells whose cost an extra center $c_\ell$ lowers};
		\end{scope}
	\end{tikzpicture}
	\caption{The hard instance (schematic). Each cell is a copy of $\Gamma$. Under $\calC_0$, rows outside $J$ are served by their own centers $p_i$, each at distance $1$ from every point in its row. The rows in $J$ have no centers of their own; their nearest center is $c_\ast$, which is at distance $u_0$ from every point. The extra centers $c_\ell$ lower the cost of the selected cells (orange) only.
		Removing outliers lowers the cost under both center sets by the same amount.}
	\label{fig:hard-instance}
\end{figure}
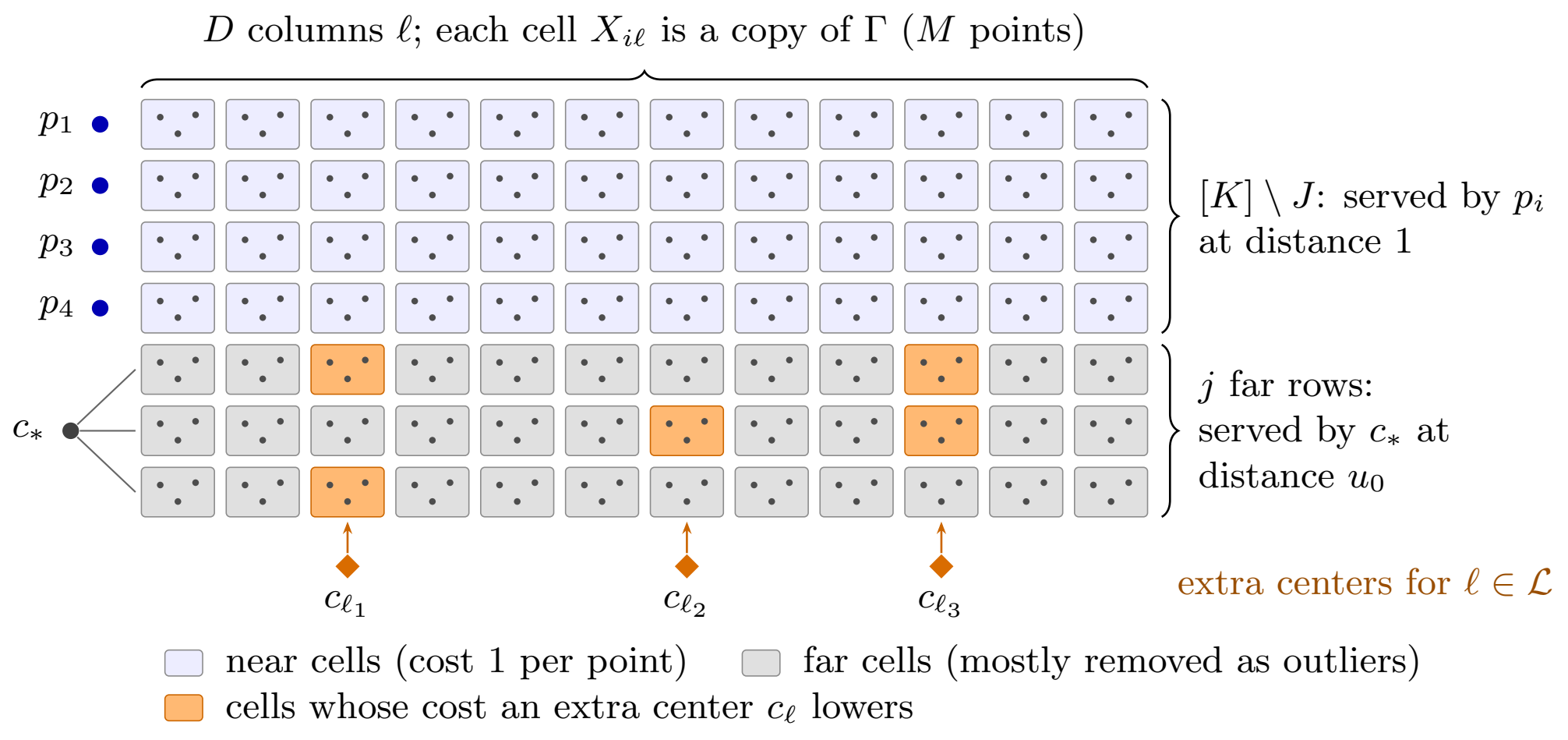

%% file: related.tex
\subsection{Related Work}

Coresets have also been studied for capacitated clustering, a variant of $(k,z)$-clustering in which each center has a capacity that limits the number of points assigned to it. Huang et al.~\cite{HJV19} constructed coresets for fair Euclidean $k$-median and $k$-means of size exponential in the dimension. Their coresets work for clustering with assignment constraints, which includes capacitated clustering as a special case. Cohen-Addad and Li~\cite{cohen2022fixed} designed a capacitated coreset of size $\mathrm{poly}(k\epsilon^{-1}\log n)$. Braverman et al.~\cite{braverman2022power} gave the first capacitated coreset of size $\mathrm{poly}(k\epsilon^{-1})$. A series of subsequent works improved this bound~\cite{huang2025coresets,in2026new,prabhu2027coresets}.
Very recently, Prabhu et al.~\cite{prabu2027price} have established the optimal coreset size for capacitated clustering in Euclidean space.

%% file: ring.tex
\section{Upper Bounds}\label{sec:upper_bounds}
Let $Y=Y_1\cup\cdots\cup Y_g$ be an $(r,k)$-regular instance, with $|Y|=n$ and $1\leq g\leq k$. Let $n_i:=|Y_i|$ and recall that $n_i\in [n/(2g),2n/g]$. Take $s$ to be a multiple of $g$, rounding up if necessary. From each $Y_i$ we draw $s/g$ points independently and uniformly with replacement, with all draws independent across clusters. Each draw from $Y_i$ has weight $gn_i/s$. The resulting weighted set $S$ assigns to each sampled point the sum of the weights of its draws. Thus $S_i:=S\cap Y_i$ has total weight $n_i$.

Recall $Q_z(k,\eps)$ from Theorem~\ref{thm:main} and let

	\[
		B_z(t,\eps) = 
		\begin{cases}
			\min\{t^{z/2},\eps^{-z}\}, & 1\leq z\leq 2, \\
			\min\{t\eps^{2-z},\eps^{-z}\}, & z>2.
		\end{cases}
	\]

Assume that $0<\eps,\gamma<1/2$. Let $L_{\eps,\gamma}=\log(k/(\eps\gamma))$.

\begin{lemma}  \label{lemma:uniform1} 
	If $s\geq C_z\gamma^{-2}L_{\eps,\gamma}^5 Q_z(k,\eps)$, then with probability at least $1-\gamma$, for every $u>0$, $J\subseteq [g]$ and $\calC\subset\mathbb{R}^d,|\calC|=k$,
	\[
		|H_{u,1}(Y_J,\calC)-H_{u,1}(S_J,\calC)|\leq \eps \cost(Y_J,\calC)+\eps n r^z.
	\]
\end{lemma}

\begin{lemma} \label{lemma:uniform2}
	Let $1\leq t\leq k$. If $s\geq C_z\gamma^{-2} L_{\eps,\gamma}^5 k\eps^{-2} B_z(t,\eps)$, then with probability at least $1-\gamma$, for every $\calC\subset\mathbb R^d$ with $|\calC|=k$, every $J\subseteq[g]$ with $|J|\leq t$ and every $u > 0$ such that for every $i\in J$,
	\begin{equation}
		u\leq \left(1+\frac{\eps}{2}\right)\max_{x\in Y_i} \dist^z(x,\calC) \label{eqn:uniform2:ubound},
	\end{equation}
	the following holds for either choice of $E\in\{Y,S\}$, with $E'$ denoting the other set:
	\begin{equation}
		H_{u,1+\eps}(E_J,\calC)-H_{u,1}(E'_J,\calC)\leq \eps n r^z. \label{eqn:uniform2}
	\end{equation}
\end{lemma}

\begin{lemma}\label{lem:regular_instance_annuli}
	Let $1\leq t\leq k$. If $n_i>2m/t$ for every $i\in[g]$ and $$
s\geq C_z\gamma^{-2}L_{\eps,\gamma}^5\big(Q_z(k,\eps)+k\eps^{-2}B_z(t,\eps)\big),
$$ then with probability at least $1-\gamma$, for every $\calC\subset\mathbb{R}^d,|\calC|=k$ and every real $0\leq b\leq \min\{m,n\}$, we have
\begin{eqnarray}
|\cost[b](S,\calC)-\cost[b](Y,\calC)|\leq \eps(\cost[b](Y,\calC)+nr^z). \label{eqn:sampling:coreset}
\end{eqnarray}
If $t=1$ and $n_i>2m$ for every $i\in[g]$, the term involving $B_z(t,\eps)$ can be omitted in the assumption on $s$. If $t=k$, the assumption $n_i>2m/t$ can be dropped, provided the term involving $B_z(t,\eps)$ is kept in the assumption on $s$.
\end{lemma}
\begin{proof}
Take the accuracy parameter $\eta=c_z\eps$ and failure probability $\gamma/2$ in Lemmata~\ref{lemma:uniform1} and~\ref{lemma:uniform2}. This changes both sample bounds by at most a constant factor depending on $z$. Condition on the event that both conclusions hold, which has probability at least $1-\gamma$.

Fix $\calC$ and $b$. Choose $E\in\{Y,S\}$ to maximize $\cost[b](E,\calC)$ and let $E'$ denote the other set. Fix an optimal removal from $E$ for $\cost[b](E,\calC)$ and let $u$ be the largest value of $\dist^z(x,\calC)$ among the remaining points, taking $u=0$ if no weight remains. Then
\begin{gather*}
\cost[b](E,\calC)=(n-b)u-H_{u,1}(E,\calC)\\
\shortintertext{and}
(1+\eta)\cdot \cost[b](E',\calC)\geq (n-b)u-H_{u,1+\eta}(E',\calC).
\end{gather*}
Subtracting gives 
\begin{equation}\label{eqn:E-E'_cost_aux}
\cost[b](E,\calC)-(1+\eta)\cdot \cost[b](E',\calC)\leq H_{u,1+\eta}(E',\calC)-H_{u,1}(E,\calC).
\end{equation}
If $u=0$, the right-hand side vanishes. For $u>0$, let $G$ consist of all indices $i$ for which $E_i$ retains at least half its weight and let $B=[g]\setminus G$. Then
\begin{equation}\label{eqn:cost_expand_aux}
H_{u,1+\eta}(E',\calC)-H_{u,1}(E,\calC)=H_{u,1+\eta}(E'_G,\calC)-H_{u,1}(E_G,\calC)+ H_{u,1+\eta}(E'_B,\calC)-H_{u,1}(E_B,\calC).
\end{equation}
By Lemma~\ref{lemma:uniform1}, we know that 
\[
H_{u,1+\eta}(E'_G,\calC)-H_{u,1}(E_G,\calC)\leq H_{u,1}(E'_G,\calC)-H_{u,1}(E_G,\calC)\leq \eta(\cost(Y_G,\calC)+n r^z).
\]
Since each cluster in $G$ retains at least half its weight after removal, the generalized triangle inequality gives
$$
\cost(Y_G,\calC)\lesssim_z \cost[b](E,\calC) + n r^z.
$$
Hence
\begin{eqnarray}
H_{u,1+\eta}(E'_G,\calC)-H_{u,1}(E_G,\calC)\lesssim_z \eta(\cost[b](E,\calC)+n r^z). \label{eqn1:sampling}
\end{eqnarray}

It remains to bound $H_{u,1+\eta}(E'_B,\calC)-H_{u,1}(E_B,\calC)$. Since $n_i/2>m/t$, we have $|B|<t$. Meanwhile, since each cluster in $B$ contains at least one outlier, we have $\max_{x\in E_i} \dist^z(x,\calC)\geq u$ for every $i\in B$. Applying Lemma~\ref{lemma:uniform2} with $u$, $B$ and $(E',E)$ gives
\begin{eqnarray}
H_{u,1+\eta}(E'_B,\calC)-H_{u,1}(E_B,\calC)\leq \eta n r^z. \label{eqn2:sampling}
\end{eqnarray}
Plugging \eqref{eqn1:sampling} and \eqref{eqn2:sampling} into \eqref{eqn:cost_expand_aux} and using $\cost[b](E',\calC)\leq\cost[b](E,\calC)$, we see from \eqref{eqn:E-E'_cost_aux} that
\[
\Delta := \cost[b](E,\calC) - \cost[b](E',\calC) \leq C_z\eta(\cost[b](E,\calC)+nr^z).
\]
Since $\cost[b](E,\calC)\leq\cost[b](Y,\calC)+\Delta$, choosing $c_z$ so that $C_z\eta\leq\eps/2$ and rearranging gives \eqref{eqn:sampling:coreset}.

If $t=1$ and $n_i>2m$ for every $i\in[g]$, then $B$ is empty and only Lemma~\ref{lemma:uniform1} is needed. If $t=k$, then $|B|\leq g\leq k$ regardless of the cluster sizes, so we may apply Lemma~\ref{lemma:uniform2} when the bound on $s$ includes the term involving $B_z(t,\eps)$.
\end{proof}

\begin{lemma}[Variation of {\cite[Lemmata~E.2--E.4]{HJLW26}}] \label{lemma:innerG} There exists a polynomial-time algorithm that, given any $(r,k)$-instance $G=G_1\cup\cdots\cup G_g$ with $G_i\subset B(a_i,r)$, constructs a coreset $S$ of size $\tilde{O}(Q_z(k,\eps))$ that, with probability at least $0.99$, satisfies the following bound for every $\calC\subset \mathbb{R}^d$ with $|\calC|=k$ and every $0\leq b\leq \min\{m,|G|\}$:
\[
|\cost[b](S,\calC)-\cost[b](G,\calC)|\leq \eps(\cost[b](G,\calC)+\cost(G,A))+C_z m r^z\eps^{1-z},
\]
where $A=\{a_1,\dots,a_g\}$. Furthermore, on the same event, $w_S(S)=|G|$.
\end{lemma}

\begin{proof}
	Let $\eta$ be the internal accuracy parameter for the coresets in~\cite{HJLW26}. By the proofs of Lemmata~E.3 and~E.4 of \cite{HJLW26}, the total failure probability is at most $\eta^z/k + 1/(10\log(k/\eta))$. Taking $\eta = c_z\eps$ for sufficiently small $c_z>0$ makes the failure probability at most $0.01$ and increases the sample size by only a constant factor.
\end{proof}

\subsection{Proof of Theorem~\ref{thm:main}}
	Assume that $|X| > m > 0$. By \cite[Section~4]{FKRS18}, we can compute a set $A\subseteq X$ of at most $2k$ centers with $\Phi:=\cost^{(m)}(X,A)\lesssim_z\min_{|\calC|=k}\cost^{(m)}(X,\calC)$. Restricting the candidate centers to $X$ costs at most a factor $2^z$, by moving each optimal center to its nearest assigned inlier. Take the $m$ points farthest from $A$ as the reference outliers, breaking ties arbitrarily. The remaining points are the reference inliers and have total cost $\Phi$ at $A$. We apply Lemmata~\ref{lem:regular_instance_annuli} and~\ref{lemma:innerG} with $k'=\max\{k,|A|\}\leq 2k$ in place of $k$, padding queries with distant centers if necessary. If $\Phi=0$, all $|X|-m$ inliers belong to $A$, so $|X|\leq m+|A|\leq m + 2k$ and we can take $X$ itself to be the coreset.

Assume that $\Phi>0$. As in \cite[Lemma~D.4]{HJLW26}, let $r_{\mathrm{out}}=(\Phi/m)^{1/z}$ and $r_{\mathrm{in}}=\eps r_{\mathrm{out}}$.

\paragraph{Handling the inner and outer groups.} The outer group consists of the $m$ reference outliers and every remaining point $x$ with $\dist(x,A)>r_{\mathrm{out}}$. Add all points in the outer group to $S$. There are at most $2m$ such points since the reference inliers have total cost $\Phi$. Let the inner group $G$ consist of the remaining points with $\dist(x,A)<r_{\mathrm{in}}$. If $G$ is nonempty, apply Lemma~\ref{lemma:innerG} to obtain $S_{\mathrm{in}}$ of size $\tilde{O}(Q_z(k,\eps))$. For every $\calC$ with $|\calC|=k$ and every $0\leq b\leq\min\{m,|G|\}$, we have
\begin{align*}
|\cost[b](S_{\mathrm{in}},\calC)-\cost[b](G,\calC)|&\leq \eps(\cost[b](G,\calC)+\cost(G,A))+C_z m r_{\mathrm{in}}^z\eps^{1-z}\\
&\lesssim_z \eps(\cost[b](G,\calC)+\Phi).
\end{align*}

\paragraph{Handling the main group.} Set $L_{\mathrm{ann}}=\lceil\log_2(r_{\mathrm{out}}/r_{\mathrm{in}})\rceil$ and $r_i=2^i r_{\mathrm{in}}$ for $i=0,\dots,L_{\mathrm{ann}}-1$. Partition the remaining points into annuli $R_i$ with $\dist(x,A)\in[r_i,2r_i)$, using $[r_{L_{\mathrm{ann}}-1},r_{\mathrm{out}}]$ for the last annulus. Then $\sum_i |R_i|r_i^z\leq\Phi$. Each $R_i$ is a $(2r_i,k')$-instance. Partition it into the Voronoi cells $R_{ij}$ of $A$.
Replace each cell $R_{ij}$ with $0<|R_{ij}|<\eps^z|R_i|/k'$ by a single point $p_{ij}\in R_{ij}$ of weight $|R_{ij}|$ as in \cite[proof of Lemma~D.2]{HJLW26}. For $x\in R_{ij}$, we have $\dist(x,p_{ij})\leq 4r_i$, so \cite[Lemma~D.5]{HJLW26} bounds the change in its cost by $\eps\dist^z(x,\calC)+C_z\eps^{1-z}r_i^z$. Any removal from $R_{ij}$ induces a removal of the same weight at $p_{ij}$, and any removal from $p_{ij}$ can be distributed among the points of $R_{ij}$. Summing the pointwise bound over the retained weight gives relative error $O(\eps)$ and additive error $O_z(\eps |R_i|r_i^z)$ on these cells, since their total weight is at most $\eps^z|R_i|$.
Group the remaining cells by $\lfloor\log_2 |R_{ij}|\rfloor$. This gives $O(\log(k/\eps))$ regular instances per annulus and $O(\log(1/\eps)\log(k/\eps))$ in total. If there are $N_{\mathrm{reg}}>0$ nonempty regular instances, use $\gamma=1/(20N_{\mathrm{reg}})$ in each application of Lemma~\ref{lem:regular_instance_annuli}.

\paragraph{Handling an $(r,k')$-regular instance.} Fix an $(r,k')$-regular instance $Y=Y_1\cup\cdots\cup Y_g$ from the preceding construction and let $1\leq t\leq k'$ be a parameter to be fixed later. Add each cluster $Y_i$ with at most $2m/t$ points directly to the coreset $S$. If any clusters remain, they lie in one dyadic size class, so their sizes differ by at most a factor of two. Their union is therefore $(r,k')$-regular, and we apply Lemma~\ref{lem:regular_instance_annuli} to it.

By a union bound, all applications of Lemma~\ref{lem:regular_instance_annuli} succeed with probability at least $0.95$.
Lemma~\ref{lemma:innerG} succeeds with probability at least $0.99$. Thus all the bounds hold simultaneously with probability at least $0.9$, and we condition on this event.

Since each part retains its total weight, minimizing over allocations of $b$ to the parts gives a global bound with the local additive errors summed, as in \cite[Fact~2.2]{HJLW26}. Their sum is
\[
	O_z\mleft(\eps\Phi+\eps\sum_i |R_i|r_i^z\mright) = O_z(\eps\Phi) = O_z\mleft(\eps\cost[b](X,\calC)\mright),
\]
where the last bound uses $b\leq m$. Rescaling $\eps$ gives the desired relative error.

\paragraph{Coreset size.} Apart from $O(m)+\tilde{O}(Q_z(k,\eps))$ points, the clusters of size at most $2m/t$ contribute $\tilde{O}(km/t)$ points in total, since $k'\leq 2k$.
For each regular instance in the main group, Lemma~\ref{lem:regular_instance_annuli}
bounds the sample size for the remaining clusters by
$\tilde{O}(Q_z(k',\eps)+k'\eps^{-2}B_z(t,\eps))$.
Summing over the $O(\log(1/\eps)\log(k/\eps))$ regular instances gives an additional
$\tilde{O}(\min\{\psi(t),k\eps^{-z-2}\})$ points beyond the vanilla bound, where
\[
\psi(t) = \begin{cases}
			kt^{z/2}\eps^{-2}, & 1\leq z\leq 2, \\
			kt\eps^{-z}, & z>2.
		\end{cases}
\]
For $t=1$, Lemma~\ref{lem:regular_instance_annuli} allows us to omit the term involving $B_z$, so the additional size is $\tilde{O}(km)$. For $t=k'$, we may instead apply the lemma to all of $Y$, regardless of the cluster sizes. The additional size is then $\tilde{O}(\min\{\psi(k'), k\eps^{-z-2}\})$.

Let $t^*$ denote the positive solution of $\psi(t)=km/t$, that is,
\[
t^*=\begin{cases} 
(m\eps^2)^{2/(z+2)}& 1\leq z\leq 2,\\
(m\eps^z)^{1/2}& z>2.
\end{cases}
\] 
If $1<t^*<k'$, take $t=\lceil t^*\rceil$. Then $km/t\leq km/t^*$ and,
since $t\leq 2t^*$, we have $\psi(t)\leq 2\psi(t^*)=2km/t^*$.
If $t^*\leq 1$, take $t=1$ and use $km\leq km/t^*$.
If $t^*\geq k'$, take $t=k'$ and use $\psi(k')\leq\psi(t^*)=km/t^*$.
Thus the additional size is $\tilde O(km/t^*)$ in all three cases.
Taking the minimum with the bounds $\tilde O(km)$ and
$\tilde O(k\eps^{-z-2})$ yields, up to logarithmic factors,
an upper bound of
\[
\begin{cases}
	\min\{km,\; km^{z/(z+2)}\eps^{-4/(z+2)},\;
	k\eps^{-z-2}\}, & 1\leq z\leq 2,\\
	\min\{km,\; k\sqrt{m}\,\eps^{-z/2},\;
	k\eps^{-z-2}\}, & z>2
\end{cases}
\]
on the additional coreset size. This is at most $F_z(k,m,\eps)$, with equality for $z>2$. For $1\leq z\leq 2$, omitting $km$ leaves the minimum unchanged when $m\geq\eps^{-2}$, since the middle term is then at most $km$. When $m<\eps^{-2}$, we have $F_z(k,m,\eps)\leq k\eps^{-2}\leq Q_z(k,\eps)$, so omitting $km$ changes the total bound by at most a constant factor.
Now we can conclude that the total coreset size is
\[
	O(m)+\tilde{O}(Q_z(k,\eps)+F_z(k,m,\eps)),
\]
as claimed.

%% file: sampling.tex
\subsection{Proof of Lemma~\ref{lemma:uniform1}}\label{sec:proof_uniform1}

If $r=0$, each cluster is a single point, so $S_J=Y_J$ as weighted sets for every $J$ and the claim follows. For $r>0$, it suffices to prove
\begin{equation}\label{eqn:uniform1}
\Pr\left\{\sup_{u,J,\calC}\frac{|H_{u,1}(Y_J,\calC)-H_{u,1}(S_J,\calC)|}{\cost(Y_J,\calC)+nr^z}>\eps\right\}\leq\gamma,
\end{equation}
where the supremum is taken over $u>0$, $J\subseteq[g]$ and sets $\calC\subset\mathbb R^d$ of $k$ centers.

\paragraph{Distance bands.} Following \cite[Section~8.2]{HJLW26}, we divide the indices into bands according to $d_i:=\dist(a_i,\calC)$. Let $T=\lceil\log_2(C_z/\eps)\rceil$, $\calI_0(\calC)=\{i\in[g]\mid d_i<r\}$ and $\calI_l(\calC)=\{i\in[g]\mid 2^{l-1}r\leq d_i<2^lr\}$ for $l\in[T]$. For $d_i\geq 2^Tr$, the distances on $Y_i$ lie in $[d_i-r,d_i+r]$. Since $h_u$ is $1$-Lipschitz and $S_i$ and $Y_i$ have the same total weight $n_i$, their ReLU sums differ by at most $n_i((d_i+r)^z-(d_i-r)^z) = O_z(r/d_i)\cost(Y_i,\calC)$, where we used the mean-value theorem and the fact that $\cost(Y_i,\calC)\geq n_i(d_i-r)^z$. Taking $C_z$ to be sufficiently large ensures that the total error from the clusters in $J$ with $d_i\geq 2^T r$ is at most $(\eps/2)\cost(Y_J,\calC)$.

For the remaining clusters, it suffices to prove for each $l\in \{0,\dots,T\}$ that
$$
\E \sup_{\calC}\sup_{u>0,J\subseteq \mathcal{I}_l(\calC)} \bigg|\frac{H_{u,1}(Y_J,\calC)-H_{u,1}(S_J,\calC)}{\cost(Y_J,\calC)+nr^z}\bigg|\leq \frac{\gamma\eps}{2(T+1)}.
$$ Indeed, restricting $J$ to $J\cap\calI_l(\calC)$ can only decrease the denominator. The sum of the $T+1$ suprema therefore bounds the normalized error from these bands. Its expectation is at most $\gamma\eps/2$, so Markov's inequality at threshold $\eps/2$ gives \eqref{eqn:uniform1}.

In the remainder of the proof, we fix $l\in\{0,\dots,T\}$ and let $\rho=2^l$.

\paragraph{Centering.} Let $g_{\calC,u,i}(x)=h_u(\dist^z(x,\calC))-h_u(\dist^z(a_i,\calC))$. For $i\in\calI_l(\calC)$ and $x\in Y_i$, the Lipschitz property of $h_u$ gives $|g_{\calC,u,i}(x)|\leq |\dist^z(x,\calC)-d_i^z|\leq zr(d_i+r)^{z-1}\lesssim_z\rho^{z-1}r^z$. As our sampling keeps the total weight on each $Y_i$, we have $$
H_{u,1}(Y_J,\calC)-H_{u,1}(S_J,\calC)=\sum_{i\in J}\left(\sum_{x\in Y_i}g_{\calC,u,i}(x)-\sum_{x\in S_i}w_S(x)g_{\calC,u,i}(x)\right).
$$ By standard symmetrization techniques, it suffices to prove that
\begin{equation}\label{eqn:simple_target_after_symmetrization}
\E \sup_{\calC}\sup_{u>0,J\subseteq \mathcal{I}_l(\calC)} \bigg|\frac{\sum_{i\in J}(gn_i/s)\sum_{a=1}^{s/g}\xi_{ia}g_{\calC,u,i}(x_{ia})}{\cost(Y_J,\calC)+nr^z}\bigg|\leq \frac{\gamma\eps}{4(T+1)},
\end{equation}
where $x_{ia}$ is the $a$-th draw from $Y_i$ and the $\xi_{ia}$ are independent Rademacher variables.

\begin{lemma}[Adaptation of {\cite[Lemma~C.1]{HJLW26}}] \label{lemma:uniform1:net}
Fix $r>0$, an integer $1\leq k'\leq k$ and centers $a_1,\dots,a_{k'}\in\mathbb R^d$. Let $P$ be a finite nonempty subset of $\bigcup_{i=1}^{k'} B(a_i,r)$. For $\rho\geq 1$ and $\alpha\in(0,1)$, consider all sets $\calC\subseteq\mathbb R^d$ of $k$ centers satisfying
$$
\rho r/4\leq r+\dist(a_i,\calC)\leq 4\rho r \qquad\text{for every }i\in[k'].
$$
The set of distance vectors $(\dist(x,\calC))_{x\in P}$ over all such $\calC$ admits an $(\alpha r)$-net $N_\alpha\subseteq\mathbb R^P$ in the $\ell_\infty$-norm with
$$
\log |N_\alpha|\lesssim k\alpha^{-2}\min\{k',\rho^2\}\log(2k|P|\rho/\alpha).
$$
\end{lemma}
\begin{proof}
Assign each point of $P$ to one ball $B(a_i,r)$ containing it, and let $X_i$ consist of the points assigned to $B(a_i,r)$. For a fixed query $\calC$, Huang et al.~\cite[Section~8.2]{HJLW26} divide the nonempty clusters into bands according to the ratio $\dist(a_i,\calC)/\Delta_i$, where $\Delta_i = (1/|X_i|)\sum_{x\in X_i}\|x-a_i\|_2$.
By adding points in orthogonal directions, we may ensure that $|X_i| = 2|P|$ and $\sum_{x\in X_i}\|x-a_i\|_2 = 2|P|r$. Thus $\Delta_i = r$ for every cluster and all clusters lie in the same cost band.

The net construction underlying \cite[Lemma~C.1]{HJLW26} applies to distance vectors on any fixed set of at most $\eta^{-1}$ points. We use $P$ as this set and take $\eta=(16k|P|\rho)^{-1}$.
For fixed $\calC$, let $J_l$ consist of the indices $i\in[k']$ for which $X_i$ has dyadic distance level $l$ in the sense of \cite[Section~8.2]{HJLW26}. 
Since $\Delta_i=r$, every $i\in J_l$ satisfies $2^l\asymp 1+\dist(a_i,\calC)/r\asymp\rho$, so only $O(1)$ bands $J_l$ are nonempty. For each such band, there are at most $k|J_l|\leq kk'$ pairs of a query center and a cluster in the band, so its interaction class $\psi$ satisfies $2^\psi\lesssim kk'$.

Fix the bands and their interaction classes. For each $J_l\neq\emptyset$, let $P_l=P\cap\bigcup_{i\in J_l}X_i$. Applying \cite[Lemma~C.1]{HJLW26} with parameter $\alpha/2$ gives a net $N_l$ of the distance vectors restricted to $P_l$. For every query $\calC$ with these bands and classes, there is a vector $v\in N_l$ such that
\[
	|v(x)-\dist(x,\calC)|\leq \frac{\alpha}{2}(\|x-a_i\|_2+\Delta_i)\leq\alpha r
	\qquad\text{for all }i\in J_l,\ x\in P\cap X_i.
\]
The last inequality uses $\|x-a_i\|_2\leq r$ for every original point $x\in P\cap X_i$. Again, by \cite[Lemma~C.1]{HJLW26},
\[
	\log|N_l|\lesssim \min\{2^\psi+k\alpha^{-2},k2^{2l}\alpha^{-2}\}\log(2k|P|\rho/\alpha)
	\lesssim k\alpha^{-2}\min\{k',\rho^2\}\log(2k|P|\rho/\alpha),
\]
where we used $k'+\alpha^{-2}\leq 2k'\alpha^{-2}$. For each $J_l$, take the union of its nets over all interaction classes and denote this union again by $N_l$. Since $2^\psi\lesssim kk'\leq k^2$, there are only $O(\log(2k))$ classes, so the union still satisfies the same log size bound.

For fixed bands $(J_l)_l$, the sets $P_l$ partition $P$. Choosing one $v_l\in N_l$ on each $P_l$ gives an $\alpha r$-net on $P$ whose logarithmic size is at most $\sum_{l:J_l\ne\emptyset}\log|N_l|$, which is a sum of $O(1)$ terms. Each of the $k'$ clusters lies in $O(1)$ possible bands, so taking the union over all possible $(J_l)_l$ adds only $\log(O(1)^{k'})=O(k')$ to the logarithmic size.
\end{proof}

\paragraph{Telescoping.}
Condition on the sampled points $x_{ia}$. Fix a nonempty $J\subseteq\calI_l(\calC)$ with $|J|=j$ and let $h_\ast = \lceil\log_2(C_z(T+1)/(\gamma\eps))\rceil$.
Let $P_J=S_J\cup\{a_i:i\in J\}$. Since $\rho r/2\leq r+d_i\leq 2\rho r$ for $i\in J$, Lemma~\ref{lemma:uniform1:net} applies on $P_J$ with $k'=|J|=j$. For each $h\in [h_\ast]$, let $b_h\in \R^{P_J}$ be a $(2^{-h}r)$-net point for $(\dist(x,\calC))_{x\in P_J}$. All these distances lie in $[0,2\rho r]$, so clipping $b_h$ to this interval preserves the approximation bound.
Let $v_0$ be the zero vector. For $h\in [h_\ast]$, $i\in J$ and $x\in S_i$, let $v_h(x)=h_u(b_h(x)^z)-h_u(b_h(a_i)^z)$. Here we suppress the dependence of $v_h$ on $u$.

Since $h_u$ is $1$-Lipschitz, the mean-value theorem gives, for each $h\in [h_\ast]$, $i\in J$ and $x\in S_i$,
\begin{equation}\label{eqn:approx_v_h}
|g_{\calC,u,i}(x)-v_h(x)|\lesssim_z 2^{-h}\rho^{z-1}r^z.
\end{equation}

By regularity and the definition of the band, the denominator satisfies
$\cost(Y_J,\calC)+nr^z\asymp_z B:=nr^z(1+jg^{-1}\rho^z)$.
For every $i\in J$, we write on $S_i$
\[
g_{\calC,u,i}=g_{\calC,u,i}-v_{h_\ast}+\sum_{h=1}^{h_\ast}(v_h-v_{h-1}).
\]
The approximation bound \eqref{eqn:approx_v_h} gives
$\|v_h-v_{h-1}\|_\infty\lesssim_z 2^{-h}\rho^{z-1}r^z$ for $h\geq2$.
For $h=1$, the same bound follows from $v_0=0$, \eqref{eqn:approx_v_h} and the centering bound $|g_{\calC,u,i}(x)|\lesssim_z\rho^{z-1}r^z$.

For each level $h\in [h_\ast]$, let
\begin{gather*}
	Z_h = \frac{1}{B}\sum_{i\in J}\frac{gn_i}{s}\sum_{a=1}^{s/g}\xi_{ia}(v_h(x_{ia})-v_{h-1}(x_{ia}))
\intertext{and define the remainder}
Z_{\textrm{rem}} = \frac{1}{B}\sum_{i\in J}\frac{gn_i}{s}\sum_{a=1}^{s/g}\xi_{ia}(g_{\calC,u,i}(x_{ia})-v_{h_\ast}(x_{ia})).
\end{gather*}
By independence of the Rademacher signs and the regularity bound $gn_i/s\leq 2n/s$,
\begin{align*}
	\Var_\xi(Z_h)
&= \frac{1}{B^2}\sum_{i\in J}\left(\frac{gn_i}{s}\right)^2
\sum_{a=1}^{s/g}\bigl(v_h(x_{ia})-v_{h-1}(x_{ia})\bigr)^2\\
&\lesssim \frac{n^2}{B^2 s^2} \frac{js}{g} \|v_h-v_{h-1}\|_\infty^2\\
&\lesssim_z \frac{2^{-2h}\rho^{2z-2}r^{2z}n^2j}{gB^2s}\\
&=\frac{2^{-2h}jg^{-1}\rho^{2z-2}}{s(1+jg^{-1}\rho^z)^2}.
\end{align*}
We next bound $\E \sup_{\calC, u, J} |Z_h|$. Let $L_s = L_{\eps,\gamma}+\log(2s)$.
For fixed $J$, $|P_J|\leq s+j\leq 2s$, so Lemma~\ref{lemma:uniform1:net} bounds the logarithm of the number of the pairs $(b_h, b_{h-1})$ by $O_z(k2^{2h}\min\{j,\rho^2\}L_s)$. For each pair, $Z_h$ is piecewise linear in $u$ with $O(s)$ breakpoints and is constant beyond the last one, so for $\sup_u |Z_h|$, it suffices to consider the breakpoints and $u=0$.

For fixed $J$, we count only pairs $(b_h,b_{h-1})$ arising from the same $\calC$ (for $h=1$, only $b_1$ matters). Let $N_J$ be the number of choices of such a pair and a threshold among its breakpoints and $0$. The $O(s)$ choices of threshold add only $O(\log(2s))$ to the logarithm of the count.
Let $\sigma_j^2$ denote the variance bound above for $|J|=j$. For each of the $N_J$ choices, the Rademacher tail bound gives $\Pr_\xi(|Z_h|>t)\leq 2\exp(-t^2/(2\sigma_j^2))$ for $t>0$. Taking a union bound over these choices and the at most $2^g$ choices of $J$ and then integrating the tail bound gives
\begin{align*}
	\E_\xi\sup_{\calC,u,J}|Z_h|
	&\lesssim \max_{J\ne\emptyset}\sigma_j\sqrt{\log(2^{g+1}N_J)}\\
	&\lesssim_z \max_{J\ne\emptyset}\sigma_j\sqrt{k2^{2h}\min\{j,\rho^2\}L_s} & (g\leq k)\\
	&\lesssim_z
	\sqrt{\frac{kL_s}{s}
		\max_{1\leq j\leq g}
		\frac{(j/g)\rho^{2z-2}\min\{j,\rho^2\}}
		{(1+(j/g)\rho^z)^2}}.
\end{align*}
To bound the fraction inside $\max_j$, observe that
\[
	\frac{(j/g)\rho^{2z-2}\min\{j,\rho^2\}}{(1+(j/g)\rho^z)^2}
	\leq \rho^{z-2}\frac{\min\{j,\rho^2\}}{1+(j/g)\rho^z}
	\leq \min\{g\rho^{-2},\rho^z\}
	\leq \min\{k\rho^{-2},\rho^z\}.
\]
For every $\rho>0$, $\min\{k\rho^{-2},\rho^z\}\leq k^{z/(z+2)}$, since the two terms are equal at $\rho=k^{1/(z+2)}$. For the bands under consideration, we also have $\rho^z\lesssim_z\eps^{-z}$.
Therefore,
\begin{equation}\label{eqn:single_band_Z_h_supremum}
	\E\sup_{\calC,u,J}|Z_h| \lesssim_z \sqrt{\frac{kL_s}{s}\min\{k^{z/(z+2)},\eps^{-z}\}} = \eps\sqrt{\frac{L_sQ_z(k,\eps)}{s}}.
\end{equation}
For the remainder, it follows from the triangle inequality and \eqref{eqn:approx_v_h} (at $h=h_\ast$) that
\[
	|Z_{\textrm{rem}}|
	\lesssim_z \frac{2^{-h_\ast}\rho^{z-1}r^z}{B}\sum_{i\in J}n_i
	\lesssim_z 2^{-h_\ast}\frac{(j/g)\rho^{z-1}}{1+(j/g)\rho^z}
	\leq 2^{-h_\ast}
	\leq \frac{\gamma\eps}{C_z(T+1)},
\]
where we used $\sum_{i\in J}n_i\lesssim nj/g$ by regularity, $\rho\geq 1$ and our choice of $h_\ast$.

Summing over the $h_\ast$ levels and adding the remainder, we obtain
\[
	\sum_{h=1}^{h_\ast} \E\sup_{\calC,u,J}|Z_h| + \E\sup_{\calC,u,J}|Z_{\mathrm{rem}}|
	\lesssim_z h_\ast\eps\sqrt{\frac{L_sQ_z(k,\eps)}{s}} + \frac{\gamma\eps}{C_z(T+1)}.
\]
Let $s_0=C_z\gamma^{-2}L_{\eps,\gamma}^5Q_z(k,\eps)$. At $s=s_0$, we have $L_s \lesssim_z L_{\eps,\gamma}$. Also note that $L_s/s$ decreases with $s$. Hence, for $s\geq s_0$,
\[
	h_\ast\eps\sqrt{\frac{L_sQ_z(k,\eps)}{s}} \lesssim_z \frac{h_\ast\gamma\eps}{L_{\eps,\gamma}^2} \lesssim_z \frac{\gamma\eps}{T+1}.
\]
Taking the constants in $h_\ast$ and $s_0$ sufficiently large gives the required bound $\gamma\eps/(4(T+1))$ for \eqref{eqn:simple_target_after_symmetrization}.

\subsection{Proof of Lemma~\ref{lemma:uniform2}}

Let $L_s=L_{\eps,\gamma}+\log(2s)$ and let $s_0=C_z\gamma^{-2} L_{\eps,\gamma}^5 k\eps^{-2} B_z(t,\eps)$ denote the sample size threshold.

\paragraph{Active clusters.} Fix a center set $\calC$ and a threshold $u$. Call a cluster $Y_i$ active with respect to $\calC$ and $u$ if $H_{u,1+\eps}(Y_i,\calC)>0$ and inactive otherwise. On an inactive cluster, the ReLU term $(u-(1+\eps)\dist^z(x,\calC))_+$ vanishes at every point, so its contribution to either difference in (\ref{eqn:uniform2}) is nonpositive. Thus it suffices to prove $(\ref{eqn:uniform2})$ assuming that $Y_i$ is active for every $i\in J$. We call such a set $J$ active. We also call $J=\emptyset$ active so that the suprema in \eqref{eqn:large_u_target} and \eqref{eqn:large_u_symmetrization} below are nonnegative.

\begin{lemma} \label{lemma:active:diam}
Let $i\in J$. If the cluster $Y_i$ is active and \eqref{eqn:uniform2:ubound} holds for $i$, then $\max_{x\in Y_i}\dist(x,\calC)\lesssim_z r/\eps$. Therefore $u\lesssim_z (r/\eps)^z$ for every nonempty active $J$ satisfying (\ref{eqn:uniform2:ubound}).
\end{lemma}
\begin{proof}
Let $d_i=\min_{x\in Y_i} \dist(x,\calC)$ and $D_i=\max_{x\in Y_i} \dist(x,\calC)$. Since $Y_i$ is active and satisfies (\ref{eqn:uniform2:ubound}), we have
$$
(1+\eps)d_i^z<u\leq (1+\eps/2)D_i^z.
$$
Hence
$$
\bigg(\frac{1+\eps}{1+\eps/2}\bigg)^{1/z} d_i\leq D_i\leq d_i+2r.
$$
Note that $$\bigg(\frac{1+\eps}{1+\eps/2}\bigg)^{1/z}=\bigg(1+\frac{\eps}{2+\eps}\bigg)^{1/z}=1+\Theta_z(\eps).
$$ Thus $d_i\lesssim_z r/\eps$ and $D_i\leq d_i+2r\lesssim_z r/\eps$. Therefore $u\leq (1+\eps/2)D_i^z\lesssim_z (r/\eps)^z$.
\end{proof}

\paragraph{Small $u$.} Suppose $u^{1/z}\leq C_z r$. Since $H_{u,1+\epsilon}(E,\calC)\leq H_{u,1}(E,\calC)$ for either $E = Y_J$ or $E=S_J$, we have
\begin{multline*}
\max\bigg\{H_{u,1+\eps}(Y_J,\calC)-H_{u,1}(S_J,\calC),H_{u,1+\eps}(S_J,\calC)-H_{u,1}(Y_J,\calC)\bigg\} \\
\leq |H_{u,1}(Y_J,\calC) - H_{u,1}(S_J,\calC)|.
\end{multline*}
For each active cluster $Y_i$, $\dist(a_i,\calC)\leq u^{1/z}+r\leq(C_z+1)r$. 
Thus only $O_z(1)$ bands occur, and $\cost(Y_J,\calC)+nr^z\asymp_z nr^z$. 
We apply the argument from the proof of Lemma~\ref{lemma:uniform1} to these $O_z(1)$ bands. Since the band parameter $\rho$ of Section~\ref{sec:proof_uniform1} satisfies $\rho\leq 2(C_z+1)=O_z(1)$ on these bands, the bound $\min\{k\rho^{-2},\rho^z\} \leq \rho^z = O_z(1)$ replaces the factor $\min\{k^{z/(z+2)}, \eps^{-z}\}$ in \eqref{eqn:single_band_Z_h_supremum}. Together with $B_z(t,\eps) \geq 1$ and the assumption on $s$, we obtain that
\[
	\E\sup_{\calC,u,J} \bigl|H_{u,1}(Y_J,\calC)-H_{u,1}(S_J,\calC)\bigr| \leq \frac{\gamma\eps nr^z}{2},
\]
where the supremum is over all triples $(\calC,u,J)$ with $\calC\subseteq\mathbb R^d$, $|\calC|=k$, $0<u^{1/z}\leq C_zr$, and $J\subseteq[g]$ active with $|J|\leq t$, such that \eqref{eqn:uniform2:ubound} holds for every $i\in J$.

\paragraph{Large $u$.} We assume $u^{1/z}>C_zr$. By Lemma~\ref{lemma:active:diam}, $u\lesssim_z (r/\eps)^z$ for nonempty active $J$. We group the queries according to the dyadic interval $[\rho r,2\rho r)$ containing $u^{1/z}$, where $\rho$ is a power of two.
Since $C_zr<u^{1/z}\lesssim_z r/\eps$ whenever $J\neq\emptyset$, only $T_u=O_z(\log(1/\eps))$ bands contain a query with $J\neq\emptyset$.
For $J=\emptyset$ both sums on the left-hand side of (\ref{eqn:uniform2}) are zero. Thus the large-$u$ bound is immediate if $T_u=0$. Otherwise we consider these $T_u$ bands and allow $J=\emptyset$ in each.

\begin{definition} For $\rho=2^l$, we define $Q_{\rho}$ to be the set of all triples $q=(\calC,u,J)$ with $\calC\subset \mathbb{R}^d, |\calC|=k$, $u^{1/z}\in [\rho r, 2\rho r)$, $J\subseteq [g],|J|\leq t$, $u^{1/z}>C_zr$ and $J$ active, such that for every $i\in J$, $$
u\leq (1+\eps/2)\max_{x\in Y_i} \dist^z(x,\calC).
$$
\end{definition}

Fix $q=(\calC,u,J)\in Q_\rho$ and $i\in J$. Let $d_i$ and $D_i$ denote the minimum and maximum of $\dist(x,\calC)$ on $Y_i$, as in the proof of Lemma~\ref{lemma:active:diam}. Since $Y_i$ is active, $d_i<u^{1/z}$. By (\ref{eqn:uniform2:ubound}), $D_i\geq u^{1/z}/(1+\eps/2)^{1/z}$. Since $D_i-d_i\leq 2r$ and $u^{1/z}>C_zr$, we have $\dist(x,\calC)\asymp \rho r$ on $Y_i$. In the following, we set $M_{\rho}=\rho^{z-1}r^z$, $K_{\rho}=k\cdot\min\{t,\rho^2\}$, $W=n/s$ and $\beta=c_z\eps\rho^zr^z$, with $c_z$ small enough that $\beta\leq M_{\rho}$ and $\beta\leq\eps\dist^z(x,\calC)$ for every $x\in Y_J$.

\begin{definition}
For a query $q =(\calC,u,J)\in Q_{\rho}$, define $v_q(x)=h_u(\dist^z(x,\calC))$ on $Y_J$ and $v_q(x)=0$ elsewhere. For two weighted datasets $E,T\subseteq Y$, define
$$
D_q(E,T)=\sum_{x\in E_J} w_E(x)(v_q(x)-\beta)_+-\sum_{x\in T_J}w_T(x)v_q(x).
$$
\end{definition}

\begin{lemma} \label{lemma:uniform2:inq1} For every $x\in Y_J$,
$v_q(x)\lesssim_z M_{\rho}$.  For every weighted dataset $E\subseteq Y_J$, $H_{u,1+\eps}(E,\calC)\leq \sum_{x\in E} w_E(x)\cdot (v_q(x)-\beta)_+$.
\end{lemma}

\begin{proof}
    Fix $i\in J$ and $x\in Y_i$. Then
    $$
    v_q(x)=(u-\dist^z(x,\calC))_+\leq (1+\eps/2) D_i^z-d_i^z\leq 2zr D_i^{z-1}+\eps D_i^z/2\lesssim_z (1+\eps\rho)M_{\rho}\lesssim_z M_{\rho}.
    $$
    Moreover, since $\beta\leq \eps \dist^z(x,\calC)$, we have $\big(u-(1+\eps)\dist^z(x,\calC)\big)_+\leq (v_q(x)-\beta)_+$. Summing over $E$ gives $H_{u,1+\eps}(E,\calC)\leq \sum_{x\in E} w_E(x)\cdot (v_q(x)-\beta)_+$.
\end{proof}

\paragraph{Symmetrization.} By Lemma~\ref{lemma:uniform2:inq1}, to prove (\ref{eqn:uniform2}) it suffices to show
\begin{equation}\label{eqn:large_u_target}
\E_S \sup_{q\in Q_{\rho}} \max\{D_q(S,Y),D_q(Y,S)\}\leq \frac{\gamma\eps n r^z}{2T_u}
\end{equation}
for each of the $T_u$ bands. Since $J=\emptyset$ is allowed, each supremum is nonnegative. We sum over the bands and include the small thresholds before applying Markov's inequality.

Let $S'$ denote an independent copy of $S$. Recall that our sampling is unbiased, so $\E_{S'}\big[D_q(S,S')\mid S\big]=D_q(S,Y)$ and $\E_{S'}\big[D_q(S',S)\mid S\big]=D_q(Y,S)$. By Jensen's inequality, 
\begin{equation}\label{eqn:large_u_symmetrization}
	\begin{aligned}
	\E_S \sup_{q\in Q_{\rho}} \max\{D_q(S,Y),D_q(Y,S)\} &\leq \E_{S,S'} \sup_{q\in Q_{\rho}} \max\{D_q(S,S'),D_q(S',S)\} \\
	&\leq 2\E \sup_{q\in Q_{\rho}} D_q(S,S').
	\end{aligned}
\end{equation}
The second inequality uses the exchangeability of $S$ and $S'$ and the fact that the suprema are nonnegative.

\paragraph{Discretization.} For $\alpha\in [\beta/4,M_{\rho}]$, call $N_\alpha\subseteq Q_\rho$ an $\alpha$-net if, for every $q\in Q_\rho$, there exists $q'\in N_\alpha$ such that $|v_q(x)-v_{q'}(x)|\leq\alpha$ for all $x\in S\cup S'$.
\begin{lemma} \label{lem:uniform2:net}
For every $\alpha\in [\beta/4,M_{\rho}]$, there exists an $\alpha$-net $N_{\alpha}\subseteq Q_\rho$ with
$$
\log |N_{\alpha}|\lesssim_z K_{\rho}M_{\rho}^2\alpha^{-2}L_s.
$$
\end{lemma}
\begin{proof} For $q=(\calC,u,J)$, recall that $v_q(x)=(u-\dist^z(x,\calC))_+$ on $Y_J$ and vanishes elsewhere. Fix a nonempty $J$. By \eqref{eqn:uniform2:ubound} and activity, for every $i\in J$ we have
\[
\rho r/4\leq u^{1/z}(1+\eps/2)^{-1/z}\leq r+\dist(a_i,\calC)\leq u^{1/z}+2r\leq 4\rho r.
\]
For $0<c'_z<1$, Lemma~\ref{lemma:uniform1:net} with $k'=|J|$ gives a $(c'_z\alpha r/M_\rho)$-net $\calN_J\subseteq\mathbb R^{S_J\cup S'_J}$ for the distance vectors. For each $q=(\calC,u,J)$, choose $b_q\in\calN_J$ such that
\[
\max_{x\in S_J\cup S'_J}|b_q(x)-\dist(x,\calC)|\le c'_z\alpha r/M_\rho
\]
and define a label for $q$ to be the tuple $(J,b_q,\lfloor u/(c'_z\alpha)\rfloor)$. Queries in $Q_\rho$ with the same label form a class.
Since $(\rho r)^z\leq u<(2\rho r)^z$ and $(\rho r)^z=\rho M_\rho$, there are at most $|\calN_J|\cdot O_z(\rho M_\rho/\alpha)$ classes for a fixed $J$.
If $q=(\calC,u,J)$ and $q'=(\calC',u',J)$ lie in the same class, then $|u-u'|<c'_z\alpha$ and $|\dist(x,\calC)-\dist(x,\calC')|\leq 2c'_z\alpha r/M_\rho$ for every $x\in S_J\cup S'_J$. Since these distances are $O(\rho r)$ on $Y_J$, the mean-value theorem gives $|\dist^z(x,\calC)-\dist^z(x,\calC')|=O_z(c'_z\alpha)$.
Choose one query from each nonempty class and let $q_\alpha$ denote the chosen query in the class containing $q$. All choices, including those of $b_q$, depend on the samples only through $S\cup S'$.
Taking $c'_z$ small enough ensures that $|v_q(x)-v_{q_\alpha}(x)|\leq\alpha$ for every $x\in S\cup S'$.
Since $\alpha\geq\beta/4$, we have $\rho M_\rho/\alpha\leq 4\rho M_\rho/\beta=4/(c_z\eps)$. Using $|J|\leq t$ and $|S_J\cup S'_J|\leq 2s$ in Lemma~\ref{lemma:uniform1:net} gives
\[
\log|\calN_J|\lesssim_z K_\rho M_\rho^2\alpha^{-2}\log\!\left(\frac{4ks\rho M_\rho}{c'_z\alpha}\right)
\lesssim_z K_\rho M_\rho^2\alpha^{-2}L_s.
\]
For fixed $J$, the last entry $\lfloor u/(c'_z\alpha)\rfloor$ of a label takes $O_z(\rho M_\rho/\alpha)=O_z(1/\eps)$ values, so it adds only $O_z(\log(1/\eps))$ to the log size.
For $J=\emptyset$, one query suffices since $v_q=0$. Taking the union over $J$ adds at most $g\log 2$ to the log size. All these contributions are absorbed in the displayed bound since $L_s\geq\log(1/\eps)$, $M_\rho/\alpha\geq1$ and $K_\rho\geq k\geq g$.
\end{proof}

\paragraph{Bounding $D_q(S,S')$.} Set $\alpha=\beta/4$ and define $l_q(x)=(v_{q_{\alpha}}(x)-\beta/4)_+$, $A_q=\{x\mid l_q(x)\ge \beta/2\}$ and $g_q(x)=(l_q(x)-\beta/4)\mathbf{1}_{A_q}(x)$.

\begin{lemma}\label{lemma:Dq}
For a query $q=(\calC,u,J)$, the following inequalities hold for every $x\in S\cup S'$:
\begin{gather*}
(v_q(x)-\beta)_+\leq g_q(x)-\frac{\beta}{4}\mathbf{1}_{A_q}(x),
\shortintertext{and}
v_q(x)\geq g_q(x)+\frac{\beta}{4}\mathbf{1}_{A_q}(x).
\end{gather*}
\end{lemma}
\begin{proof} Since $q_{\alpha}$ is the net point assigned to $q$ and $\alpha=\beta/4$, we have $l_q(x)\leq(v_q(x)+\beta/4-\beta/4)_+= v_q(x)$ and $l_q(x)\geq (v_q(x)-\beta/4-\beta/4)_+\geq v_q(x)-\beta/2$.

If $x\in A_q$, we have $(v_q(x)-\beta)_+\leq l_q(x)-\beta/2=g_q(x)-\beta/4$ and $v_q(x)\geq l_q(x)=g_q(x)+\beta/4$.

If $x\not \in A_q$, we have $(v_q(x)-\beta)_+\leq (l_q(x)+\beta/2-\beta)_+=0$. The right-hand sides of both inequalities are also zero.
\end{proof}

Within each cluster, pair the draws from $S$ and $S'$ in sampling order. Index the pairs by $a\in[s]$, write $(x_a,x_a')$ for the $a$-th pair and let $w_a\in[W/2,2W]$ be the common weight of its two draws. By Lemma~\ref{lemma:Dq}, we have
\begin{equation}\label{eqn:D_q(S,S')_first_upper_bound}
\begin{aligned}
\E \sup_{q\in Q_{\rho}} D_q(S,S') &= \E \sup_{q\in Q_{\rho}} \sum_a  w_a \big((v_q(x_a)-\beta)_+-v_q(x_a')\big)\\
&\leq
\E \sup_{q\in Q_{\rho}} \sum_a  w_a \big(g_q(x_a)-\frac{\beta}{4}\mathbf{1}_{A_q}(x_a)-g_q(x'_a)-\frac{\beta}{4}\mathbf{1}_{A_q}(x'_a)\big)
\\
&=\E \sup_{q\in Q_{\rho}} \bigg(\sum_a \xi_a w_a \big(g_q(x_a)-g_q(x'_a)\big)-\frac{\beta}{4}\sum_a w_a\big(\mathbf{1}_{A_q}(x_a)+\mathbf{1}_{A_{q}}(x_a')\big)\bigg),
\end{aligned}
\end{equation}
where the $\xi_a$ are independent Rademacher variables. The last equality holds because swapping $x_a$ and $x'_a$ changes neither the joint distribution of the pairs nor $S \cup S'$ (hence nor the net), while it flips the sign of $g_q(x_a) - g_q(x'_a)$ for every $q$ and leaves the second sum unchanged.

Define $\calG=\{g_q\mid q\in Q_\rho\}\cup\{0\}$. For $f\in\calG$, write $\|f\|_\infty=\max_{x\in S\cup S'}|f(x)|$ and let $Z_f=\sum_a \xi_a w_a \big(f(x_a)-f(x'_a)\big)$.
For any $A\subseteq Y$, let $\mu(A)=\sum_a w_a\big(\mathbf{1}_A(x_a)+\mathbf{1}_A(x_a')\big)$. It follows immediately from \eqref{eqn:D_q(S,S')_first_upper_bound} that
\[
\E \sup_{q\in Q_{\rho}} D_q(S,S')\leq \E\sup_{f\in\calG}\big(Z_f-\tfrac{\beta}{4}\mu(\supp(f))\big).
\]

We next bound the right-hand side.

\begin{lemma} \label{lemma:uniform2:zg}
	It holds that
    \[
    \E\sup_{f\in\calG}\big(Z_f-\tfrac{\beta}{4}\mu(\supp(f))\big)\lesssim_z \beta^{-1}WM_{\rho}^2K_{\rho} L_s\log^2(1/\eps)\log(2s).
    \]
\end{lemma}

Condition on the sampled pairs $(x_a,x_a')_{a=1}^s$. For an arbitrary $P\geq W$, define $\calG_P=\{f\in\calG\mid\mu(\supp(f))\leq P\}$.

\begin{lemma} \label{lemma:uniform2:Z2}
	It holds that
    \[
	\E_{\xi} \max_{f\in\calG_P} Z_f^2\lesssim_z WPM_{\rho}^2K_{\rho} L_s\log^2(1/\eps).
	\]
\end{lemma}

\begin{proof}
	For each $\theta\in[c_z\beta,C_zM_{\rho}]$, we choose a $\theta$-net $\calN_\theta\subseteq\calG_P$ and bound $\log|\calN_\theta|$. First, consider the case where $\theta \geq 4\beta$. Let $\delta = \min\{\theta/4, M_{\rho}\}$. Group the functions $g_q\in\calG_P$ according to $q_\delta$ and choose one function from each group. Together with the zero function, these functions form a $\theta$-net: by Lemma~\ref{lemma:Dq}, $\|g_q-v_q\|_\infty\leq\beta$, so two functions in the same group are at distance at most $2\beta+2\delta\leq \theta$. Thus $|\calN_\theta|\leq |N_\delta|+1$.
	Now consider the case where $\theta < 4\beta$. Take $\calN_\theta = \calG_P$. Since each $g_q$ is determined by $q_{\beta/4}$, we have $|\calN_\theta|\leq |N_{\beta/4}|+1$. In both cases, Lemma~\ref{lem:uniform2:net} gives	$\log|\calN_\theta|\lesssim_z K_{\rho} M_{\rho}^2 \theta^{-2}L_s$.
	
	Let $h_0=\lceil\log_2\beta\rceil$ and $h_1=\lceil\log_2 M_{\rho}\rceil$. For each $f\in\calG_P$ and $h_0\leq h\leq h_1$, let $f_h$ denote its net point in $\calN_{2^h}$.
	Since $2^{h_0}<2\beta$, we may take $f_{h_0}=f$. Set $f_{h_1+1}=0$. Write $Z_{h,f}=Z_{f_h}-Z_{f_{h+1}}$, so that $Z_f=\sum_{h=h_0}^{h_1} Z_{h,f}$. We have $\|f_h-f_{h+1}\|_\infty\lesssim_z 2^h$, using $\|f_{h_1}\|_\infty\lesssim_z M_{\rho}\leq 2^{h_1}$ at $h=h_1$. Thus
	\begin{align*}
	\Var(Z_{h,f})
	&= \sum_a w_a^2(f_h(x_a)-f_h(x'_a)-(f_{h+1}(x_a)-f_{h+1}(x_a')))^2\\
	&\lesssim W\sum_a w_a \big((f_h(x_a)-f_{h+1}(x_a))^2+(f_h(x_a')-f_{h+1}(x_a'))^2\big)\\
	&\lesssim_z W 2^{2h}\big(\mu(\supp(f_h))+\mu(\supp(f_{h+1}))\big)\\
	&\lesssim WP 2^{2h}.
	\end{align*}
	The logarithm of the number of pairs $(f_h, f_{h+1})$ is $O_z(2^{-2h} K_{\rho} M_{\rho}^2L_s)$. Hence,
	\[
		\E \max_{f\in\calG_P} |Z_{h,f}|^2 \lesssim_z WPM_{\rho}^2K_{\rho}L_s.
	\]
	Since $M_{\rho}/\beta=(c_z\eps\rho)^{-1}$ and $\rho\geq1$, there are $h_1-h_0+1=O_z(\log(1/\eps))$ levels.
	Summing over these levels yields
	\[
		\left(\E \max_{f\in\calG_P}Z_f^2\right)^{1/2}
		\leq \sum_{h=h_0}^{h_1}\left(\E_\xi\max_{f\in\calG_P}|Z_{h,f}|^2\right)^{1/2}
		\lesssim_z \log\frac{1}{\eps}\cdot \sqrt{WPM_{\rho}^2K_{\rho}L_s}.
		\qedhere
	\]
\end{proof}

\begin{proof}[Proof of Lemma~\ref{lemma:uniform2:zg}]
If $\mu(\supp(f))\leq W$, then
\[
Z_f-\tfrac{\beta}{4}\mu(\supp(f))\leq |Z_f|\leq \|f\|_\infty\mu(\supp(f))\lesssim_z M_{\rho}W\leq \beta^{-1}WM_{\rho}^2K_{\rho}.
\]
Group the remaining functions according to $P/2<\mu(\supp(f))\leq P$, where $P=2^hW$ and $1\leq h\leq\lceil\log_2(2s)\rceil$. These ranges suffice since $\mu(Y)=2n=2sW$. On each range,
\[
\big(Z_f-\tfrac{\beta}{4}\mu(\supp(f))\big)_+\leq (Z_f-\beta P/8)_+\leq \frac{2Z_f^2}{\beta P}.
\]
By Lemma~\ref{lemma:uniform2:Z2}, the contribution of each range is at most
\[
\frac{2}{\beta P}\E_\xi\max_{f\in\calG_P} Z_f^2
\lesssim_z \beta^{-1}WM_{\rho}^2K_{\rho} L_s\log^2(1/\eps).
\]
Summing over the $O(\log(2s))$ ranges, adding the bound for $\mu(\supp(f))\leq W$ and taking expectation over the pairs proves the claim.
\end{proof}

\paragraph{Conclusion.} For each band corresponding to $u^{1/z}\in [\rho r,2\rho r)$, we have shown that
\begin{align*}
\E \sup_{q\in Q_{\rho}} D_q(S,S') &\lesssim_z \frac{n r^z}{s\eps}\, k\rho^{z-2}\min\{t,\rho^2\}\,L_s\log^2\left(\frac{1}{\eps}\right)\log(2s) \\
&\leq \frac{n r^z}{s\eps}\, k\rho^{z-2}\min\{t,\rho^2\}\,L_s^2 \log^2\left(\frac{1}{\eps}\right),
\end{align*}
where we used the fact that $\log(2s)\leq L_s$.
Observing that
\[
\max_{1\leq \rho \leq C_z/\eps} \rho^{z-2}\min\{t,\rho^2\}\asymp_z B_z(t,\eps)
\] 
and that $L_s^2/s$ is decreasing when $s\geq s_0$, we have
\[
	\E \sup_{q\in Q_{\rho}} D_q(S,S') \lesssim_z \frac{n r^z}{{s_0}\eps}\, kB_z(t,\eps)\, L_{s_0}^2 \log^2\left(\frac{1}{\eps}\right)
	\lesssim_z \frac{n r^z}{s_0 \eps}\, kB_z(t,\eps) L_{\eps,\gamma}^4,
\]
where we used the facts that $\log(1/\eps)\lesssim L_{\eps,\gamma}$ and $L_{s_0}\lesssim_z L_{\eps,\gamma}$. Substituting the value of $s_0$ yields
\[
	\E \sup_{q\in Q_{\rho}} D_q(S,S') \lesssim_z \frac{\gamma^2 \eps n r^z}{L_{\eps,\gamma}} \leq \frac{\gamma \eps n r^z}{4T_u}
\]
by choosing $C_z$ to be sufficiently large.
Equation~\eqref{eqn:large_u_symmetrization} now gives \eqref{eqn:large_u_target}.

\subsection{Proof of Theorem~\ref{thm:geb}}
Since projecting a center onto $B(0,1)$ cannot increase its distance to any $x\in B(0,1)$, the infimum in $\mathcal E_n^{(\nu)}$ may be taken over $\calC\subseteq B(0,1)$. For each $\nu\in[0,1]$, the optimality of $\calC_X^*$ gives
\[
	\calE_n^{(\nu)} \leq 2\E_{X\sim\calD^n}
	\sup_{\calC} \left|\widehat{\calR}^{(\nu)}(X,\calC) - \calR^{(\nu)}(\calD,\calC)
	\right|.
\]
Let $H_{u,1}(\calD,\calC) = \E_{x\sim\calD} h_u(\operatorname{dist}^z(x,\calC))$. Note that $H_{u,1}(\calD,\calC) = \int_0^1(u-q_{\calC}(v))_+\,dv$, so
\[
	(1-\nu)u - H_{u,1}(\calD,\calC) \leq \int_0^{1-\nu} \min\{u, q_{\calC}(v)\}dv \leq \calR^{(\nu)}(\calD,\calC).
\] 
Since $q_{\calC}$ is nondecreasing, both inequalities are equalities at $u = \sup\{q_{\calC}(v): 0 < v < 1-\nu\}$ (with $\sup\emptyset = 0$), so $\calR^{(\nu)}(\calD,\calC)=\sup_{u\geq 0}\{(1-\nu)u-H_{u,1}(\calD,\calC)\}$. By Lemma~\ref{lem:dual} with $\lambda=1$ and $m=\nu n$, the same identity holds for $\hat\calR^{(\nu)}(X,\calC)$ with $\frac{1}{n} H_{u,1}(X,\calC)$ in place of $H_{u,1}(\calD,\calC)$. Therefore
\[
\left|\hat\calR^{(\nu)}(X,\calC)-\calR^{(\nu)}(\calD,\calC)\right| \le \sup_{u\ge 0} \left|\frac{1}{n} H_{u,1}(X,\calC) - H_{u,1}(\calD,\calC)\right|,
\]
and it suffices to show
\[
\E_{X\sim \calD^n} \sup_{\calC,u} \left|\frac{1}{n}H_{u,1}(X,\calC)-H_{u,1}(\calD,\calC)\right| \lesssim_z \sqrt{\frac{k}{n}}\log^{\frac{3}{2}}(2kn).
\]
The rest of the proof follows that of Lemma~\ref{lemma:uniform1}. We list only the differences.
Let $g_{\calC,u}(x)=h_u(\dist^z(x,\calC))-h_u(\dist^{z}(0,\calC))$. By standard symmetrization, it suffices to show
\[
\frac{1}{n}\mathbb{E}\sup_{\calC,u} \left|\sum_{i=1}^n \xi_i g_{\calC,u}(x_i)\right| \lesssim_z \sqrt{\frac{k}{n}}\log^{\frac{3}{2}}(2kn).
\]
For the telescoping, apply Lemma~\ref{lemma:uniform1:net} on $\{0,x_1,\dots,x_n\}$ with $a_1=0$ and $k'=r=\rho=1$. 
At scale $\alpha = 2^{-h}$, the increment $v_h - v_{h-1}$ is determined by the pair $(b_h, b_{h-1})$ and, for the supremum over $u$, by a threshold among its $O(n)$ breakpoints and $0$. By Lemma~\ref{lemma:uniform1:net}, the logarithm of the number of such choices is $O(k\alpha^{-2}\log(2kn/\alpha))$. As in the proof of Lemma~\ref{lemma:uniform1}, $\|v_h-v_{h-1}\|_\infty\lesssim_z\alpha$, so each normalized Rademacher increment has subgaussian norm $O_z(\alpha/\sqrt n)$.
The Rademacher bound therefore gives 
\[
	\E_\xi \sup_{\calC,u} \left|\frac{1}{n}\sum_{i=1}^n \xi_i (v_h(x_i)-v_{h-1}(x_i))\right|
	\lesssim_z \frac{\alpha}{\sqrt{n}}\sqrt{k\alpha^{-2}\log\frac{2kn}{\alpha}} =  \sqrt{\frac{k}{n}\log\frac{2kn}{\alpha}}.
\]	
Summing over $O(\log n)$ dyadic scales down to $n^{-1/2}$ and adding the $O_z(n^{-1/2})$ residual yields the desired bound.

%% file: lb.tex
\section{Lower Bound}

\subsection{Setup}\label{sec:lb_setup}
Let $R = R(m,k,\eps)\geq 2z$ be a parameter to be determined and $u_0 = \sqrt{2R^2 - R + 1}$. Define $\phi_{u_0}\colon[-1,1]\to\mathbb R$ by
\[
\phi_{u_0}(t) = \frac{1}{z u_0^{z-1}}\left(u_0^z - \left(u_0^2 - \frac{u_0}{4}t_+\right)^{z/2}\right).
\]
This is a perturbation of the scaled ReLU function $t\mapsto t_+/8$. Since $u_0\geq 2z$, we have $0\leq\phi_{u_0}(t)\leq 1/4$ on $[-1,1]$.

Let $K = K(m,k,\eps)$ and $D = D(m,k,\eps)$ be positive integers to be determined. Let $e_1,\dots,e_K$ be the standard basis vectors of $\R^K$ and $f_1,\dots,f_D$ be the standard basis vectors of $\R^D$. Let $q\geq 2$ be an even integer to be specified in Lemma~\ref{lem:Y} and let $\Gamma\subseteq\bS^{q+2}$ have cardinality $M$. The hard instance is a point set $X$ in
\[
\R^K \oplus \R^D \oplus \R^{q+3}\oplus \R\oplus \R \simeq \R^{K+D+q+5}.
\]
For each $i\in [K]$, $\ell\in [D]$ and $y\in \Gamma$, define
\[
x_{i\ell y} = \left(Re_i, \frac{1}{\sqrt 2}f_\ell, \frac{1}{\sqrt 2}y, 0, 0 \right)\in \R^K \oplus \R^D \oplus \R^{q+3}\oplus \R\oplus \R.
\]
Hence the hard instance contains $|X| = KDM$ points in total. For $i\in[K]$ and $\ell\in[D]$, the cell $X_{i\ell}=\{x_{i\ell y}:y\in\Gamma\}$ contains exactly $M$ points. For $J\subseteq[K]$, let $X_J=\bigcup_{i\in J,\,\ell\in[D]}X_{i\ell}$.

\begin{lemma}\label{lem:two_center_sets}
	Let $J\subseteq [K]$ satisfy $1\leq |J| = j \leq R^2$ and let $\calL\subseteq[D]$. For each $\ell\in \calL$, choose a unit vector $v_\ell \in \bS^{q+2}$ and a subset $I_\ell\subseteq J$. If
	\[
	K-j+|\calL|+1\le k,
	\]
	then there are two sets $\calC_0$ and $\calC_1$ of exactly $k$ centers such that, for every $i\in[K]$, $\ell'\in[D]$ and $y\in \Gamma$,
	\[
	\dist^z(x_{i\ell' y},\calC_0) - \dist^z(x_{i\ell' y},\calC_1) =
	\begin{cases}
		z u_0^{z-1}\phi_{u_0}(\langle y,v_{\ell'}\rangle),& \ell'\in \calL\text{ and }i\in I_{\ell'},\\
		0,&\text{otherwise.}
	\end{cases}
	\]
	In particular, $\cost(X,\calC_1)\le\cost(X,\calC_0)$.
\end{lemma}
\begin{proof}
	Note that $\norm{x_{i\ell y}}_2 = \sqrt{R^2+1}$ for all $i,\ell,y$. For each $i\in [K]$, define a center point
	\[
	p_i = (Re_i,0,0,0,0).
	\]
	Then
	\[
	\|x_{i\ell' y} - p_{i'}\|_2 =
	\begin{cases}
		1, & i = i', \\
		\sqrt{2R^2 + 1}, & i\neq i'.
	\end{cases}
	\]
	Define another center point
	\[
	c_\ast = (0, 0, 0, \sqrt{R^2-R}, 0).
	\]
	Then
	\[
	\|x_{i\ell' y}-c_\ast\|_2^2 = (R^2+1) + (R^2-R) = 2R^2 - R + 1 = u_0^2.
	\]
	For each $\ell\in \calL$ define a center point
	\[
	c_\ell = \left(\frac{1}{4}\sum_{h\in I_\ell} e_h, \frac{R}{2\sqrt{2}}f_\ell, \frac{u_0}{4\sqrt{2}}v_\ell, \eta_\ell, 0	\right),
	\]
	where
	\[
	\eta_\ell^2 = R^2 - \frac{|I_\ell|}{16} - \frac{R^2}{8} - \frac{u_0^2}{32}.
	\]
	This choice is well defined because $|I_\ell| \le |J| = j \le R^2$ and $u_0^2 < 2R^2$ give
	\[
	\eta_\ell^2\geq R^2 - \frac{R^2}{16} - \frac{R^2}{8} - \frac{2R^2}{32} = \frac{3}{4}R^2 > 0.
	\]
	Next we compute $\norm{x_{i\ell' y} - c_\ell}_2^2$.
	\begin{align*}
		\|x_{i\ell' y} - c_\ell\|_2^2 
		& = \|x_{i\ell' y}\|_2^2 + \|c_\ell\|_2^2 - 2\langle x_{i\ell' y}, c_\ell\rangle\\
		& = (R^2 + 1) + R^2 - 2\left( \frac{R}{4}\mathbf{1}_{\{i\in I_\ell\}} + \frac{R}{4}\mathbf{1}_{\{\ell = \ell'\}} + \frac{u_0}{8}\inner{y}{v_\ell} \right) \\
		&= u_0^2 - \frac{u_0}{4}\inner{y}{v_\ell} + \frac{R}{2}\left(2-\mathbf{1}_{\{i\in I_\ell\}} - \mathbf{1}_{\{\ell'=\ell\}}\right).
	\end{align*}
	If either indicator is zero, the squared distance exceeds $u_0^2$, since $\inner{y}{v_\ell}\leq 1$ and $R/2 > u_0/4$.
	
	Now we define $\calC_0$ and $\calC_1$ as
	\[
	\calC_0 = \{c_\ast\}\cup\{p_i: i\not\in J\},\qquad \calC_1 = \calC_0 \cup \{c_\ell: \ell\in \calL\}.
	\]
	Then
	\[
	\dist(x_{i\ell' y}, \calC_0) = 
	\begin{cases}
		u_0, & i\in J, \\
		1, & i \not\in J.
	\end{cases}
	\]
	If $i\in I_\ell$ and $\ell'=\ell$, then $i\in J$ and the cost of $x_{i\ell' y}$ under $\calC_0$ is $u_0^z$. Adding $c_\ell$ to $\calC_0$ changes this cost to
	\[
	\min\{u_0^2,u_0^2-(u_0/4)\inner{y}{v_\ell}\}^{z/2}
	= u_0^z - z u_0^{z-1}\phi_{u_0}(\inner{y}{v_\ell}).
	\]
	Thus
	\[
	\dist^z(x_{i\ell' y}, \calC_1) =
	\begin{cases}
		u_0^z - z u_0^{z-1}\phi_{u_0}(\inner{y}{v_{\ell'}}), & \ell'\in \calL\text{ and }i\in I_{\ell'},\\
		u_0^z, & i\in J\text{ and the preceding condition fails},\\
		1, & i\notin J.
	\end{cases}
	\]
	If necessary, pad each of $\calC_0$ and $\calC_1$ to exactly $k$ centers with distinct points of the form $(0,0,0,0,t)$ with $|t|>R$. Their squared distance from every point of $X$ is $R^2+1+t^2>u_0^2$, so they preserve the distances to both center sets.
\end{proof}

%

\subsection{Point Set \texorpdfstring{$\Gamma$}{Gamma}}
\begin{lemma}\label{lem:Y}
	There are absolute constants $c,C>0$ and $0<\bar\tau\leq 1/2$ such that the following holds. Let $u_0\geq 2z$ and $\tau \in (0, \bar\tau]$. Define $N = \lceil c/(\tau^2\log^3(1/\tau))\rceil$ and $q = \min\{h\in 2\Z: h\geq 2,\ 2^h\geq 512N\}$. One can choose a set of unit directions
	\[
	V = \{v_b: b\in\{-1,1\}^q\} \subseteq \bS^{q+2}
	\]
	such that for every integer $M\ge C\tau^{-2}$, there exists a set of exactly $M$ distinct unit vectors
	\[
	\Gamma = \{y_1,\dots,y_M\} \subseteq\bS^{q+2}
	\]
	satisfying
	\[
	\frac{1}{|V|} \sum_{v\in V} \left| \sum_{y\in \Gamma}\phi_{u_0}(\langle y,v\rangle) - \sum_{y\in \Gamma'}w_y\phi_{u_0}(\langle y,v\rangle)\right| \ge M\tau
	\]
	for every subset $\Gamma'\subseteq \Gamma$ with $|\Gamma'|\le N$ and every choice of nonnegative weights $\{w_y:y\in \Gamma'\}$.
\end{lemma}

The proof of Lemma~\ref{lem:Y} is deferred to Section~\ref{sec:proof_Y}.

\subsection{Directions with Large Discrepancy}
Let $S$ be a weighted subset of $X$ and let $w_{i\ell y}=w_S(x_{i\ell y})$. For each $i\in [K]$ and $\ell\in [D]$, define the discrepancy for the cell $(i,\ell)$ in the direction $v\in \bS^{q+2}$ as
\[
\Delta_{i\ell}(v) = \sum_{y\in \Gamma} \phi_{u_0}(\inner{y}{v}) - \sum_{x_{i\ell y}\in S} w_{i\ell y} \phi_{u_0}(\inner{y}{v}).
\]
\begin{lemma}\label{lem:bad_directions}
	Let $u_0$, $\tau$, $V$, $N$ and $\Gamma$ be as in Lemma~\ref{lem:Y}. Fix a subset $J\subseteq [K]$, a subset $\calL\subseteq [D]$ and a weighted subset $S\subseteq X$ with nonnegative weights. For each $\ell\in \calL$, let $\mathcal{R}_\ell\subseteq J$ satisfy $|S\cap X_{i\ell}|\leq N$ for every $i\in\mathcal{R}_\ell$. Then there exist directions $v_\ell\in V$ and subsets $I_\ell\subseteq\mathcal{R}_\ell$ such that
	\[
	\left| \sum_{\ell\in \calL}\sum_{i\in I_\ell} \Delta_{i\ell}(v_\ell) \right| \geq \frac{1}{2}M\tau \sum_{\ell\in \calL} |\mathcal{R}_\ell|.
	\]
\end{lemma}
\begin{proof}
	For each $\ell\in \calL$, it follows from Lemma~\ref{lem:Y} that
	\[
	\frac{1}{|V|}\sum_{v\in V}\sum_{i\in\mathcal{R}_\ell} |\Delta_{i\ell}(v)| \geq M\tau|\mathcal{R}_\ell|.
	\]
	Choose $v_\ell\in V$ such that
	\[
		\sum_{i\in\mathcal{R}_\ell} |\Delta_{i\ell}(v_\ell)| \geq M\tau|\mathcal{R}_\ell|.
	\]
	Summing over $\ell$, 
	\[
		\sum_{\ell\in\calL} \sum_{i\in\mathcal{R}_\ell} |\Delta_{i\ell}(v_\ell)| \geq M\tau\sum_{\ell\in\calL}|\mathcal{R}_\ell|.
	\]
	Split the sum on the left-hand side into two sums, according to the sign of $\Delta_{i\ell}(v_\ell)$. One of them must be at least $(1/2)M\tau\sum_{\ell\in\calL}|\calR_\ell|$. Fix this sign for all $\ell\in\calL$ and let $I_\ell$ consist of the indices $i\in\mathcal{R}_\ell$ for which $\Delta_{i\ell}(v_\ell)$ has this sign. This proves the claim.
\end{proof}

\subsection{Lower Bound for Robust Clustering}\label{sec:robust_lb}
Let $C,\bar\tau$ be as in Lemma~\ref{lem:Y}, decreasing $\bar\tau$ so that $\bar\tau\leq\sqrt C$. Let $E_z=1+16(3/2)^z$ and $A_0>24E_z$.

\begin{lemma}\label{lem:construction_support}
	Let $R\geq 2z$ and let $j,K,D$ be positive integers with $j\leq\min\{R^2,K/2\}$. Suppose that $0<\eps\leq 1/(4(E_z+1))$ and $\tau=A_0\eps R\leq\bar\tau$. Let $M=\lceil C/\tau^2\rceil$ and let $N$ and $\Gamma$ be as in Lemma~\ref{lem:Y} with $u_0=\sqrt{2R^2-R+1}$. Let $X$ be the instance of Section~\ref{sec:lb_setup} with $n=KDM$ points. If $jR^z\leq D\leq 2jR^z$, $D\geq 32K$, $2K\leq k$ and $jDM\leq m\leq n$, then every robust $(\eps,m)$-coreset $S\subseteq X$ satisfies $|S|\geq KDN/4$.
\end{lemma}
\begin{proof}
	Suppose $|S|<KDN/4$. By averaging, at least $K/2$ indices $i\in[K]$ satisfy $\sum_{\ell\in[D]}|S\cap X_{i\ell}|\leq DN/2$. Since $j\leq K/2$, we can let $J$ consist of $j$ of them. Call a cell $X_{i\ell}$ sparse if $|S\cap X_{i\ell}|\leq N$. Since $|S\cap X_J|\leq jDN/2$, at least $jD/2$ of the cells $X_{i\ell}$ with $i\in J$ are sparse.
	
	Let $\calC_0$ be the center set from Lemma~\ref{lem:two_center_sets} for $J$. Recall that points of $X_J$ are at distance $u_0$ from $\calC_0$ and all other points are at distance $1$. Let $m_0=jDM$ and set the budget $b_0=m_0-16jKM$. Since $D\geq 32K$ and $m_0\leq m$, we have $0\leq b_0\leq m$.
	Since the points of $X_J$ are farthest from $\calC_0$, the $b_0$ removed points all come from $X_J$. Hence
	\[
		P_0:=\cost[b_0](X,\calC_0)=n-m_0+16u_0^zjKM\leq E_z n,
	\]
	where we used $u_0\leq 3R/2$ and $D\geq jR^z$. 
	
	Next we show that $W_J=w_S(S\cap X_J)$ is close to $m_0 = |X_J|$. Since the origin is equidistant from all points of $X$, applying the coreset guarantee at budget $b_0$ to the origin and $k-1$ sufficiently distant centers gives $|w_S(S)-n|\leq\eps(n-b_0)$. Moreover,
	\[
		\cost[b_0](S,\calC_0)=w_S(S)-b_0+(u_0^z-1)(W_J-b_0)_+.
	\]
	Comparing with $P_0=n-b_0+(u_0^z-1)(m_0-b_0)$ gives
	\[
		\left|(W_J-b_0)_+-(m_0-b_0)\right|
		\leq \frac{\eps P_0+|w_S(S)-n|}{u_0^z-1}
		\leq 4(E_z+1)\eps jKM<2jKM,
	\]
	where the second inequality uses $u_0^z-1\geq R^z/2$ and $D\leq 2jR^z$. Since $m_0-b_0=16jKM$, it follows that $W_J>b_0$ and $|W_J-m_0|<2jKM$.
	
	For each $\ell\in[D]$, let $\mathcal R_\ell$ be the set of $i\in J$ for which $X_{i\ell}$ is sparse and has coreset weight at most $8M$. At most $W_J/(8M)<(jD+2jK)/8\leq 17jD/128$ cells $X_{i\ell}$ with $i\in J$ have weight greater than $8M$. Thus $\sum_{\ell=1}^D|\mathcal R_\ell|\geq jD/3$. Let $\calL\subseteq[D]$ consist of $K$ indices $\ell$ with the largest $|\mathcal R_\ell|$. Then $\sum_{\ell\in\calL}|\mathcal R_\ell|\geq jK/3$.
	
	Lemma~\ref{lem:bad_directions} gives directions $v_\ell$ and sets $I_\ell\subseteq\mathcal R_\ell$ such that
	\begin{equation}\label{eqn:robust_cell_discrepancy}
	z u_0^{z-1}\left|\sum_{\ell\in\calL}\sum_{i\in I_\ell}\Delta_{i\ell}(v_\ell)\right|
	\geq \frac{1}{6}R^{z-1}M\tau jK
	\geq \frac{A_0}{12}\eps n.
	\end{equation}
	The last inequality uses $\tau=A_0\eps R$ and $D\leq 2jR^z$. Since $K+|\calL|=2K\leq k$, Lemma~\ref{lem:two_center_sets} gives the second center set $\calC_1$.
	
	The cells with $\ell\in\calL$ and $i\in I_\ell$ contain at most $jKM$ points of $X$ and at most $8jKM$ weight of $S$. Outside these cells, $X_J$ still contains at least $m_0-jKM>b_0$ points of $X$ and at least $W_J-8jKM>b_0$ weight of $S$, all at the maximum distance $u_0$ under both center sets. Hence removing $b_0$ units of weight at this distance lowers each of $\cost(X,\calC_0)$, $\cost(S,\calC_0)$, $\cost(X,\calC_1)$ and $\cost(S,\calC_1)$ by exactly $b_0u_0^z$.
	
	It follows from Lemma~\ref{lem:two_center_sets} that the left-hand side of \eqref{eqn:robust_cell_discrepancy} equals
	\[
	\left|\big(\cost[b_0](X,\calC_0)-\cost[b_0](S,\calC_0)\big)
	-\big(\cost[b_0](X,\calC_1)-\cost[b_0](S,\calC_1)\big)\right|.
	\]
	The coreset guarantee bounds this by $2\eps P_0\leq 2E_z\eps n$, since $\dist(x,\calC_1)\leq\dist(x,\calC_0)$ for every $x\in X$. This contradicts $A_0>24E_z$.
\end{proof}

\begin{theorem}\label{thm:global_LB}
	Let $1\leq z\leq 2$. There is a constant $\eps_0>0$ depending only on $z$ such that, for every integer $k\geq 1$, every integer $m\geq 0$ and every $\eps\in(0,\eps_0]$, there exists a finite point set $X$ in a Euclidean space with $|X| > m$ such that every robust $(\eps,m)$-coreset $S\subseteq X$ for $(k,z)$-clustering satisfies
	\[
	|S| = \Omega\!\left(m + \frac{\min\left\{k\eps^{-(z+2)},\, k^{\frac{2(z+1)}{z+2}}\eps^{-2} + k m^{\frac{z}{z+2}}\eps^{-\frac{4}{z+2}}\right\}}{\log^3(1/\eps)}\right).
	\]
\end{theorem}
\begin{proof}
	Choose $\eps_0$ small enough that $\Lambda = \log(1/\eps)\geq 1$. Let $Q=Q_z(k,\eps)$ and $F=F_z(k,m,\eps)$ as defined in Theorem~\ref{thm:main}.
	Up to constant factors, the expression inside $\Omega(\cdot)$ in the theorem is $m + (Q+F)/\Lambda^3$.
	Below we shall prove lower bounds $B_Q = c_Q Q/\Lambda^3$ and $B_F = c_F F/\Lambda^3$, where $c_Q,c_F > 0$ depend only on $z$. These bounds and the elementary lower bound $\Omega(m)$ may use different hard instances. Choosing the instance with the largest lower bound suffices, since $\max\{m,B_Q,B_F\} \geq (m + B_Q + B_F)/3$.
	If $\max\{B_Q,B_F\} > m$, the corresponding hard instance has more than $m$ points, since the full input is itself an exact coreset. Otherwise, we use an instance of $m + 1$ distinct points.
	
	Theorem~\ref{thm:vanilla_LB} gives the bound $B_Q$. Together with the elementary $\Omega(m)$ lower bound, this suffices when $F\leq A_z(m+Q)$, where $A_z$ is a sufficiently large constant depending only on $z$. Henceforth, assume that $F > A_z(m+Q)$. This forces $k\leq\eps^{-(z+2)}$, since otherwise $F\leq Q$. Thus $Q = k^{(2z+2)/(z+2)}\eps^{-2}$. Comparing $F > A_zQ$ with $F\leq km^{z/(z+2)}\eps^{-4/(z+2)}$ gives $m > A_z^{(z+2)/z}k\eps^{-2}$. In particular, $m\eps^2 > 1$, so $F\leq km$ and hence $k > A_z$.
	
	Let $A_0$, $C$ and $\bar\tau$ be as fixed at the start of Section~\ref{sec:robust_lb}. Define
	\begin{gather*}
		R = \min\left\{\left(\frac{A_0^2\eps^2m}{16C}\right)^{1/(z+2)},\,\frac{\bar\tau}{A_0\eps}\right\},
		\quad
		j = \lfloor R^2\rfloor, \\
		K = \left\lfloor\frac{k}{2}\right\rfloor,
		\quad
		D = \lceil jR^z\rceil,
		\quad
		\tau = A_0\eps R,
		\quad
		M = \left\lceil\frac{C}{\tau^2}\right\rceil,
		\quad
		m_0 = jDM,
		\quad
		n = KDM.
	\end{gather*}
	For sufficiently large $A_z$ and sufficiently small $\eps_0$, we have $R\geq 2z$ and $\tau\leq\bar\tau$. Since $R^2/2\leq j\leq R^2$, $jR^z\leq D\leq 2jR^z$ and $C\tau^{-2}\leq M\leq 2C\tau^{-2}$, we have
	\[
	m_0 \leq \frac{4Cj^2R^{z-2}}{A_0^2\eps^2} \leq \frac{4CR^{z+2}}{A_0^2\eps^2} \leq \frac{m}{4}
	\]
	and
	\[
	n \asymp_z kR^z\eps^{-2} \asymp_z \min\{km^{z/(z+2)}\eps^{-4/(z+2)},k\eps^{-(z+2)}\} = F.
	\]
	For sufficiently large $A_z$, our assumption on $F$ therefore gives
	\[
	n \geq 32m,
	\qquad
	\frac{j}{K} = \frac{m_0}{n} \leq \frac{m}{4n} \leq \frac{1}{128}.
	\]
	We also have $D\geq 32K$. If $D<32K$, then $D\leq 2R^{z+2}$ would give
	\[
	n = KDM \lesssim_z kD\eps^{-2}R^{-2}
	\lesssim kD^{z/(z+2)}\eps^{-2}
	\lesssim k^{(2z+2)/(z+2)}\eps^{-2} = Q,
	\]
	contradicting $n\asymp_z F > A_zQ$ for sufficiently large $A_z$.
	
	Thus $n>m$ and, by Lemma~\ref{lem:construction_support}, every robust $(\eps,m)$-coreset of the instance $X$ of Section~\ref{sec:lb_setup} has at least $(1/4)KDN$ points. Since $N\gtrsim \tau^{-2}/\log^3(1/\tau)$ by Lemma~\ref{lem:Y} and $M\asymp \tau^{-2}$, we have
	\[
		KDN \gtrsim \frac{n}{\log^3(1/\tau)} \gtrsim_z \frac{F}{\Lambda^3},
	\]
	where the last inequality uses $\log(1/\tau)\leq\Lambda$ since $\tau=A_0\eps R\geq\eps$. Combining this with the vanilla and elementary lower bounds proves the theorem.
\end{proof}

\begin{theorem}\label{thm:global_LB_z>2}
	Let $z>2$. There is a constant $\eps_0>0$ depending only on $z$ such that, for every integer $k\geq 1$, every integer $m\geq 0$ and every $\eps\in(0,\eps_0]$, there exists a finite point set $X$ in a Euclidean space with $|X|>m$ such that every robust $(\eps,m)$-coreset $S\subseteq X$ for $(k,z)$-clustering satisfies
	\[
	|S| = \Omega\!\left(m + \frac{\min\left\{k\eps^{-(z+2)},\, k^{\frac{2(z+1)}{z+2}}\eps^{-2} + k\min\{m,\sqrt{m}\eps^{-\frac{z}{2}}\}\right\}}{\log^3(1/\eps)}\right).
	\]
\end{theorem}
\begin{proof}
	Choose $\eps_0$ small enough that $\Lambda = \log(1/\eps)\geq 1$. Let $Q=Q_z(k,\eps)$ and $F=F_z(k,m,\eps)$ as defined in Theorem~\ref{thm:main}.
	As in the proof of Theorem~\ref{thm:global_LB}, the vanilla lower bound $\Omega(Q/\Lambda^3)$ and the elementary $\Omega(m)$ lower bound suffice when $F\leq A_z(m+Q)$, where $A_z$ is a sufficiently large constant depending only on $z$. Henceforth, assume that $F>A_z(m+Q)$. This forces $k\leq\eps^{-(z+2)}$, since otherwise $F\leq Q$. Thus $Q=k^{(2z+2)/(z+2)}\eps^{-2}$. It then suffices to prove an $\Omega(F/\Lambda^3)$ lower bound. Since $F\leq km$ and $Q\geq k\eps^{-2}$, we have $k>A_z$ and $m>A_z\eps^{-2}$.
	
	Let $A_0$, $C$ and $\bar\tau$ be as fixed at the start of Section~\ref{sec:robust_lb}. Define
	\begin{gather*}
		R = \min\left\{\left(\frac{A_0^2\eps^2m}{16C}\right)^{1/(z-2)},\,\frac{\bar\tau}{A_0\eps}\right\},
		\quad
		j_\ast = \min\left\{R^2,\,\left(\frac{A_0^2\eps^2m}{16CR^{z-2}}\right)^{1/2}\right\},  \\
		\tau = A_0\eps R,
		\quad
		M = \left\lceil \frac{C}{\tau^{2}}\right\rceil,
		\quad
		j = \lfloor j_\ast\rfloor,
		\quad
		D = \lceil jR^z\rceil,
		\quad
		K = \left\lfloor\frac{k}{2}\right\rfloor,
		\quad
		m_0 = jDM,
		\quad
		n = KDM.
	\end{gather*}
	For sufficiently large $A_z$ and sufficiently small $\eps_0$, we have $R\geq 2z$ and $\tau\leq\bar\tau$.
	Since $R^{z-2}\leq A_0^2\eps^2m/(16C)$, the definition of $j_\ast$ gives $1\leq j_\ast\leq R^2$. Hence $1\leq j\leq R^2$ and $j_\ast/2\leq j\leq j_\ast$.
	Since $jR^z\leq D\leq 2jR^z$ and $C\tau^{-2}\leq M\leq 2C\tau^{-2}$, we have
	\[
	m_0=jDM\leq\frac{4Cj^2R^{z-2}}{A_0^2\eps^2}\leq\frac m4.
	\]
	If the first term in the minimum defines $R$, then $j_\ast=j=1$ and $n\asymp_z km$. Otherwise, $R=\bar\tau/(A_0\eps)$, $\tau=\bar\tau$ and $M\asymp1$, so
	\[
	j\asymp_z\min\{\eps^{-2},\sqrt m\eps^{z/2}\},
	\qquad
	n\asymp_z k\min\{\eps^{-(z+2)},\sqrt m\eps^{-z/2}\}.
	\]
	The two terms defining $R$ agree at $m = (16C/A_0^2)(\bar\tau/A_0)^{z-2}\eps^{-z}$. In the first case, $m\lesssim_z\eps^{-z}$, so $F\asymp_z km$. In the second case, $m\gtrsim_z\eps^{-z}$, so $km$ can be omitted from the minimum defining $F$, up to constant factors. Thus, in both cases,
	\[
	n\asymp_z\min\{km,k\sqrt m\eps^{-z/2},k\eps^{-(z+2)}\}=F.
	\]
	As in the proof of Theorem~\ref{thm:global_LB}, the bounds $m_0\leq m/4$, $n\asymp_z F > A_z(m+Q)$ and $j\leq R^2$ imply that $n\geq 32m$, $j/K\leq 1/128$ and $D\geq 32K$, for sufficiently large $A_z$. Lemma~\ref{lem:construction_support} therefore gives an instance $X$ of size $n>m$ with
	\[
	|S| \geq \frac{1}{4}KDN \gtrsim \frac{n}{\log^3(1/\tau)} \gtrsim_z \frac{F}{\Lambda^3}
	\]
	for every robust $(\eps,m)$-coreset $S\subseteq X$, where we used $M\asymp\tau^{-2}$ and $\log(1/\tau)\leq\Lambda$ since $\tau=A_0\eps R\geq\eps$. This completes the proof.
\end{proof}

\subsection{Proof of Lemma~\ref{lem:Y}}\label{sec:proof_Y}
The proof is based on Boolean Fourier analysis; see, e.g., \cite[Lemma~3.1]{LWW19}. We shall prove the spectral estimates needed for $\phi_{u_0}$.

For an even integer $q\ge 2$, define
\[
\mathcal{U}_q = \{a\in\{-1,1\}^q : a_1 = 1\},\qquad H = |\mathcal{U}_q| = 2^{q-1}.
\]
For $E\subseteq [q]$, we define the Walsh character
\[
\chi_E(a) = \prod_{i\in E} a_i,\qquad a\in \mathcal{U}_q.
\]
For $t\ge 0$, let $B_t$ be the $H\times H$ matrix with entries
\[
(B_t)_{ab} = \left(\frac{|\inner{a}{b}|}{q} - t\right)_+, \qquad a,b\in\mathcal U_q.
\]
We denote by $\sigma_{\min}(T)$ the smallest singular value of a matrix $T$.

\begin{lemma}\label{lem:B_t}
	The matrices $\{B_t\}_{t\geq 0}$ have a common orthogonal eigenbasis indexed by	the subsets $E\subseteq[q]$ of even cardinality. Let $\beta_E(t)$ be the eigenvalue of $B_t$ corresponding to $E$. There is an absolute constant $c_1>0$ such that
	\[
	\sigma_{\min}(B_0)\ge c_1\frac{\sqrt H}{q^{3/2}}.
	\]
	Moreover, for every $E\subseteq[q]$ of even cardinality and every $t\ge0$,
	\[
	|\beta_E(t)|\le |\beta_E(0)|.
	\]
\end{lemma}
\begin{proof}
	For every $E\subseteq[q]$ of even cardinality, $\chi_E$ is an eigenvector of $B_t$. Indeed, for $a\in\mathcal U_q$, the change of variables $b_i\mapsto a_i b_i$ in the sum over $b\in\mathcal U_q$ gives $(B_t\chi_E)_a=\chi_E(a)(B_t\chi_E)_{\mathbf{1}}$. If $E$ and $F$ are distinct subsets of even cardinality, they differ at some coordinate $i\ne 1$. Flipping this coordinate in the sum for $\inner{\chi_E}{\chi_F}$ shows that the sum is zero. There are $2^{q-1}=H$ subsets of even cardinality, so their characters form a common orthogonal eigenbasis. The eigenvalue for $E$ is
	\[
	\beta_E(t) = \beta_E(t)\chi_E(\mathbf{1}) = (B_t \chi_E)_{\mathbf{1}} = \sum_{b\in \mathcal{U}_q}(B_t)_{\mathbf{1},b} \chi_E(b) = \sum_{b\in \mathcal{U}_q} \left( \frac{|\sum_i b_i|}{q} - t \right)_+  \chi_E(b).
	\]	
	Extend $\chi_E$ to $\{-1,1\}^q$ using the same formula. Since $|E|$ is even, $\chi_E(b) = \chi_E(-b)$ for every $b\in\{-1,1\}^q$, so
	we can write
	\[
	\beta_E(t) = \frac{1}{2} \sum_{b\in \{-1,1\}^q} \left( \frac{|\sum_i b_i|}{q} - t \right)_+  \chi_E(b).
	\]
	
	Suppose that $E\neq\emptyset$. Let $s_b = \sum_i b_i$. Using the identities
	\[
	\sum_{b\in  \{-1,1\}^q} \chi_E(b) = 0,\qquad (|s|-a)_+ = \frac{|s-a|+|s+a|}{2} - a,\quad a\geq 0,
	\]
	we can write $\beta_E(t)$ as
	\[
	\beta_E(t) 
	= \frac{1}{2q}\sum_{b\in\{-1,1\}^q} \left(\frac{|s_b - qt| + |s_b + qt|}{2} - qt\right)\chi_E(b) 
	= \frac{1}{4q}\sum_{b\in\{-1,1\}^q} \left(|s_b - qt| + |s_b + qt|\right)\chi_E(b).
	\]
	Now, substituting
	\[
	|s| = \frac2\pi\int_0^\infty\frac{1-\cos(s\theta)}{\theta^2}\,d\theta,
	\]
	we obtain that
	\begin{align*}
		\beta_E(t) 
		&= -\frac{1}{2\pi q} \int_0^\infty\frac{\sum_b \chi_E(b) (\cos((s_b - qt)\theta) + \cos((s_b + qt)\theta))}{\theta^2}\,d\theta \\
		&= -\frac{1}{\pi q} \int_0^\infty \frac{\cos(qt\theta)}{\theta^2} \left(\sum_b \chi_E(b) \cos(s_b \theta)\right) d\theta.
	\end{align*}
	Since $|E|$ is even, pairing $b$ with $-b$ shows that $\sum_b\chi_E(b)e^{i s_b\theta}$ is real. Hence,
	\[
	\sum_b \chi_E(b) \cos(s_b \theta) = \sum_b \chi_E(b) e^{i s_b \theta} = \sum_b \chi_E(b) \prod_h e^{i b_h \theta}.
	\]
	Factoring the last sum coordinatewise gives
	\[
	\sum_b \chi_E(b) \prod_h e^{i b_h \theta} 
	= \left(\prod_{h\in E} \sum_{b\in\{-1,1\}} b e^{i b \theta}\right)  \left(\prod_{h\not\in E} \sum_{b\in\{-1,1\}} e^{i b \theta}\right) 
	= 2^q(i\sin\theta)^{|E|}(\cos\theta)^{q-|E|}.
	\]
	Hence, noticing that $r = |E|$ is even,
	\begin{equation}\label{eqn:beta_E_integral}
		\beta_E(t) = (-1)^{\frac{r}{2}+1}\cdot \frac{2H}{\pi q}
		\int_0^\infty \frac{\sin^r\theta\cos^{q-r}\theta}{\theta^2}	\cos(qt\theta)\,d\theta.
	\end{equation}
	These integrals converge absolutely, since $r\geq 2$. Both $r$ and $q-r$ are
	even, so the fraction in the integrand is nonnegative. It follows that 
	$|\beta_E(t)|\leq |\beta_E(0)|$ for $E\neq\emptyset$.
	
	For the lower bound on $|\beta_E(0)|$, restrict the integral to $I = [\frac{\pi}{4} - \frac{1}{4\sqrt{q}},\frac{\pi}{4}]$ if $r-q/2 < 0$ and $I = [\frac{\pi}{4},\frac{\pi}{4} + \frac{1}{4\sqrt{q}}]$ if $r-q/2\geq 0$. Then $\tan^{r-q/2}\theta \geq 1$ on $I$ and
	\[
	\int_0^\infty \frac{\sin^r\theta\cos^{q-r}\theta}{\theta^2}\,d\theta
	\geq \int_I \frac{(\sin\theta\cos\theta)^{q/2}\tan^{r-\frac{q}{2}}\theta}{(\pi/2)^2}\,d\theta
	\gtrsim \int_I (\sin\theta\cos\theta)^{q/2}\,d\theta.
	\]
	For $\theta\in I$,
	\[
	\sin\theta\cos\theta = \frac{1}{2}\cos\left(2\theta - \frac{\pi}{2}\right) \geq \frac{1}{2}\left(1 - \frac{1}{8q}\right),
	\]
	so
	\[
	\int_I (\sin\theta\cos\theta)^{q/2}\,d\theta \gtrsim \frac{1}{2^{q/2}\sqrt{q}}.
	\]
	It follows from \eqref{eqn:beta_E_integral} that for $E\neq\emptyset$,
	\[
	|\beta_E(0)|\gtrsim \frac{H}{q\cdot 2^{q/2}\sqrt{q}} \asymp \frac{\sqrt{H}}{q^{3/2}}.
	\]
	
	Finally, we consider $E = \emptyset$. Since $\chi_{\emptyset}$ is the all-ones vector, the eigenvector equation $B_t\chi_{\emptyset} = \beta_{\emptyset}(t)\chi_{\emptyset}$ shows that every row of $B_t$ sums to $\beta_{\emptyset}(t)$. Since $0\leq (B_t)_{ab}\leq (B_0)_{ab}$, we have $0\leq\beta_{\emptyset}(t)\leq\beta_{\emptyset}(0)$. Moreover, for any nonempty $F$ of even cardinality, the triangle inequality gives
	\[
	\beta_{\emptyset}(0) = \sum_{b\in\mathcal{U}_q} (B_0)_{\mathbf{1},b}
	\geq \left| \sum_{b\in\mathcal{U}_q} (B_0)_{\mathbf{1},b}\chi_F(b) \right|
	= |\beta_F(0)| \gtrsim \frac{\sqrt{H}}{q^{3/2}}. \qedhere
	\]
\end{proof}

The next lemma concerns the matrix for $\phi_{u_0}$.
\begin{lemma}\label{lem:phi_u_kernel}
	Let $u_0\geq 2z$ and let $q\geq 2$ be even. Define a matrix $A\in \R^{H\times H}$ indexed by $\mathcal{U}_q$ by
	\[
	A_{ab} = \frac{1}{2}\left[ \phi_{u_0}\!\left(\frac{\inner{a}{b}}{2q}\right) + \phi_{u_0}\!\left(-\frac{\inner{a}{b}}{2q}\right) \right].
	\]
	Then
	\[
	\sigma_{\min}(A)\ge c\frac{\sqrt H}{q^{3/2}},
	\]
	for some absolute constant $c > 0$.
\end{lemma}
\begin{proof}
	Let $\psi_{u_0}(s) = \phi_{u_0}(s/2)/2$ for $s\in [0,1]$. Then
	$A_{ab} = \psi_{u_0}(|\inner{a}{b}|/q)$ and
	\[
	\psi_{u_0}(0)=0,\qquad	\psi_{u_0}'(s)=\frac1{32}\left(1-\frac{s}{8u_0}\right)^{z/2-1}.
	\]
	Since $u_0\geq 2z$ and $\psi_{u_0}''$ has constant sign,
	\[
	\psi_{u_0}'(0)=\frac{1}{32},\qquad
	\int_0^1 |\psi_{u_0}''(t)| \, dt
	= |\psi_{u_0}'(1)-\psi_{u_0}'(0)| < 0.04 \psi_{u_0}'(0).
	\]
	The last inequality follows from
	$|\log(\psi_{u_0}'(1)/\psi_{u_0}'(0))|\leq |z/2-1|/(8u_0-1)\leq 1/30$.
	The identity
	\[
	\psi_{u_0}(s) = \psi_{u_0}'(0)s + \int_0^1 \psi_{u_0}''(t)(s-t)_+\,dt
	\]
	gives 
	\[
	A = \psi_{u_0}'(0)B_0 + \int_0^1 \psi_{u_0}''(t) B_t\,dt.
	\]
	By Lemma~\ref{lem:B_t}, all these matrices have a common orthogonal eigenbasis indexed by subsets $E\subseteq [q]$ of even cardinality. Let $\alpha_E$ be the eigenvalue of $A$ corresponding to $\chi_E$. Then
	\[
	|\alpha_E| \geq \psi_{u_0}'(0)|\beta_E(0)| - \int_0^1 |\psi_{u_0}''(t)|\,|\beta_E(t)|\,dt
	\geq 0.96 \psi_{u_0}'(0)|\beta_E(0)|.
	\]
	The lower bound then follows from Lemma~\ref{lem:B_t}.
\end{proof}

We now construct the point set $\Gamma$. First we construct a probability measure on a finite set $Z\subseteq\bS^{q+2}$ such that every nonnegative measure supported on at most $N$ points of $Z$ has a large average absolute discrepancy over $V$. The approximating measure need not have total weight $1$. We then draw samples from the probability measure and make repeated samples distinct without changing their inner products with any direction in $V$.

Assume that $q\geq 6$ is even. In $\R^q\oplus \R\oplus \R^2$, define
\begin{gather*}
	\xi_a = \left(\frac{a}{\sqrt{2q}}, 0, \frac{1}{\sqrt{2}}, 0\right),\quad a\in \mathcal{U}_q,\\
	v_b = \left(\frac{b}{\sqrt{2q}}, \frac{1}{\sqrt{2}}, 0, 0\right),\quad b\in \{-1,1\}^q,
\end{gather*}
which are unit vectors in $\R^{q+3}$. Let $Z = \{\xi_a: a\in\mathcal{U}_q\}$ and $V = \{v_b: b\in \{-1,1\}^q\}$. Then $|Z| = H = 2^{q-1}$ and $|V| = 2^q$. For a real vector $w = (w_a)_{a\in \mathcal{U}_q}$, define
\[
F_w(v) = \sum_{a\in \mathcal{U}_q} w_a \phi_{u_0}(\inner{\xi_a}{v}).
\]
Then, for the matrix $A$ in Lemma~\ref{lem:phi_u_kernel},
\[
(Aw)_{b} = \frac{1}{2}(F_w(v_b) + F_w(v_{-b})).
\]
It follows that
\begin{equation}\label{eqn:discrepancy_in_ell_1_norm}
	\frac{1}{|V|}\sum_{v\in V} |F_w(v)| \geq \frac{1}{H}\norm{Aw}_1.
\end{equation}

\begin{lemma}\label{lem:probability_measure_Z}
	Suppose that $H \geq 256N$ and $\sigma_{\min}(A) > 0$. There exists a probability measure $\sum_{\xi_a\in Z} p_a \delta_{\xi_a}$ on $Z$ such that for every nonnegative vector $w$ with at most $N$ nonzero entries,
	\[
	\frac{1}{|V|}\sum_{v\in V}\left| \sum_{a\in\mathcal{U}_q} p_a\phi_{u_0}(\inner{\xi_a}{v}) - \sum_{a\in\mathcal{U}_q} w_a\phi_{u_0}(\inner{\xi_a}{v}) \right|\geq \frac{\sigma_{\min}(A)}{8H}.
	\]
\end{lemma}
\begin{proof}
	Let
	\[
	\mathcal{P} = \{\sgn(Aw): w\in\R^H, \, |\supp w|\leq 2N\}.
	\] 
	Here we take $\sgn(0) = 1$ and denote by $d_H$ the Hamming distance on $\{-1,1\}^H$. 
	We first seek a sign vector $\sigma\in \{-1,1\}^H$ such that for every $\omega\in \mathcal{P}$,
	\[
	d_H(\omega,\sigma) \geq \frac{H}{4}.
	\]
	
	Fix a set $I\subseteq\mathcal{U}_q$ of $2N$ column indices. For $w$ supported on $I$, each coordinate $(Aw)_b$ is a linear form in the $2N$ variables $(w_a)_{a\in I}$. By the extension of Warren's theorem to allow zeros~\cite[Corollary~2.1]{GJ95}, there are at most
	\[
	\left(\frac{8eH}{2N}\right)^{2N} = \left(\frac{4eH}{N}\right)^{2N}
	\]
	sign patterns of $\sgn(Aw)$ for $w$ supported on $I$. There are at most $\binom{H}{2N}$ such choices of $I$. Thus
	\begin{align*}
		|\mathcal{P}| &\leq \binom{H}{2N} \left(\frac{4eH}{N}\right)^{2N}\\
		&\leq \left(\frac{eH}{2N}\right)^{2N} \left(\frac{4eH}{N}\right)^{2N}\\
		&= \exp\left(4N\left(1 + \ln\frac{H}{N}\right)+2N\ln 2\right) \\
		& < \exp(H/8)
	\end{align*}
	by the assumption that $H\geq 256N$.
	Now, consider a uniformly random $\sigma\sim \{-1,1\}^H$. For a fixed $\omega\in\mathcal{P}$, $d_H(\omega,\sigma)\sim \operatorname{Bin}(H,1/2)$ and thus
	\[
		\Pr_\sigma\left\{ d_H(\omega,\sigma) < \frac{H}{4} \right\} \leq e^{-H/8}.
	\]
	By a union bound over all $\omega\in\mathcal{P}$,
	\[
		 \Pr_\sigma\left\{ \exists \omega\in\mathcal{P},\, d_H(\omega,\sigma) < \frac{H}{4} \right\} \leq |\mathcal{P}|e^{-H/8} < 1.
	\]
	Hence there exists $\sigma\in\{-1,1\}^H$ such that $d_H(\omega,\sigma)\geq H/4$ for all $\omega\in P$.
	
	We now use $\sigma$ to construct $\lambda\in\R^H$ such that $A\lambda$ is far from $Aw$ for every $(2N)$-sparse $w$.
	Let $\eta = \sigma_{\min}(A)/H$ and $\lambda = \eta A^{-1}\sigma$. Then $A\lambda = \eta\sigma$ and
	\[
	\norm{\lambda}_1 \leq \sqrt{H}\norm{\lambda}_2 \leq \frac{\sqrt{H}\eta}{\sigma_{\min}(A)}\norm{\sigma}_2 = 1.
	\]
	For any $(2N)$-sparse vector $w$, the signs of $Aw$ and $A\lambda = \eta\sigma$ differ in at least $H/4$ coordinates. In each such coordinate, the entries differ by at least $\eta$, so
	\[
	\frac{1}{H}\norm{A(\lambda-w)}_1 \geq \frac{\eta}{4}.
	\]
	Write $\lambda = \lambda_+ - (-\lambda)_+$. One of the vectors $\lambda_+$ and $(-\lambda)_+$, which we denote by $\mu$, satisfies
	\[
	\frac{1}{H}\norm{A(\mu - w)}_1 \geq \frac{\eta}{8}
	\]
	for every $N$-sparse vector $w$. Otherwise, subtracting $N$-sparse approximations to the two parts would give a $2N$-sparse vector $w$ with $\norm{A(\lambda-w)}_1<H\eta/4$ by the triangle inequality, contradicting the preceding bound. Let $p = \mu/\norm{\mu}_1$. Since $\norm{\mu}_1\leq 1$, we have
	\[
	\frac{1}{H}\norm{A(p - w)}_1 = \frac{1}{\norm{\mu}_1}\frac{1}{H}\norm{A(\mu - \norm{\mu}_1 w)}_1 \geq \frac{\eta}{8\norm{\mu}_1}\geq \frac{\eta}{8}.
	\]
	Together with \eqref{eqn:discrepancy_in_ell_1_norm}, this gives the desired inequality.
\end{proof}

Next, we convert the probability measure on $Z$ in Lemma~\ref{lem:probability_measure_Z} to an unweighted point set.

\begin{lemma}\label{lem:unweighted_conversion}
	Suppose that $p$ is a probability measure on $Z$ and
	\[
	\frac{1}{|V|}\sum_{v\in V}|F_p(v)-F_w(v)|\ge \kappa
	\]
	for every nonnegative vector $w$ with at most $N$ nonzero entries. For every integer $M\geq 1$, there is a set $\Gamma\subseteq\mathbb S^{q+2}$ of exactly $M$ distinct points such that, for every $\Gamma'\subseteq \Gamma$ with $|\Gamma'|\le N$
	and all nonnegative weights $w_y$,
	\[
	\frac1{|V|}\sum_{v\in V}
	\left|\sum_{y\in \Gamma}\phi_{u_0}(\langle y,v\rangle)
	-\sum_{y\in \Gamma'}w_y\phi_{u_0}(\langle y,v\rangle)\right|
	\ge M\left(\kappa-\frac1{4\sqrt M}\right).
	\]
\end{lemma}
\begin{proof}
	Draw $a^{(1)},\dots,a^{(M)}$ independently from $\mathcal{U}_q$ with $\Pr\{a^{(h)}=a\}=p_a$ and define
	\[
	y_h = \left(\frac{a^{(h)}}{\sqrt{2q}}, 0, \frac{\zeta_h}{\sqrt{2}}\right)\in \R^q\oplus \R\oplus \R^2,
	\]
	where $\zeta_1,\dots,\zeta_M\in\bS^1$ are distinct, so $y_1,\dots,y_M$ are distinct. Since the last two coordinates of every $v\in V$ are zero, we have
	\[
	\inner{y_h}{v} = \inner{\xi_{a^{(h)}}}{v},\qquad \forall v\in V.
	\]
	Consider
	\[
	G(v) = \frac{1}{M}\sum_{h=1}^M \phi_{u_0}(\inner{y_h}{v}),
	\]
	which is the average of $M$ i.i.d.\ random variables in $[0,1/4]$. It is clear that $\E G(v) = F_p(v)$. Hence,
	\[
	\E\left(\frac{1}{|V|}\sum_{v\in V}|G(v) - F_p(v)|\right) \leq \frac{1}{|V|} \sum_{v\in V}\sqrt{\Var(G(v))} \leq \frac{1}{4\sqrt{M}}.
	\]
	Therefore, there exists a choice of $a^{(1)},\dots,a^{(M)}$ such that
	\[
	\frac{1}{|V|}\sum_{v\in V}|G(v) - F_p(v)| \leq \frac{1}{4\sqrt{M}}.
	\]
	Fix such a choice and let
	\[
	\Gamma = \{y_1,\dots,y_M\}.
	\]
	
	Now, fix any $\Gamma'\subseteq \Gamma$ with $|\Gamma'|\leq N$ and nonnegative weights $w_y$.
	Define for each $a\in \mathcal{U}_q$,
	\[
	w'_a := \frac{1}{M}\sum_{\substack{h: y_h\in \Gamma'\\a^{(h)} = a}} w_{y_h}.
	\]
	Then $w'$ is nonnegative with at most $|\Gamma'|\leq N$ nonzero entries. Since $\inner{y_h}{v} = \inner{\xi_{a^{(h)}}}{v}$ for every $v\in V$,
	\begin{align*}
		F_{w'}(v) &= \frac{1}{M}\sum_{a\in\mathcal{U}_q} \left(\sum_{\substack{h: y_h\in \Gamma'\\a^{(h)} = a}} w_{y_h}\right) \phi_{u_0}(\inner{\xi_a}{v})\\
		&= \frac{1}{M}\sum_{a\in\mathcal{U}_q}\sum_{\substack{h: y_h\in \Gamma'\\a^{(h)} = a}} w_{y_h}\phi_{u_0}(\inner{y_h}{v})\\
		&= \frac{1}{M}\sum_{h: y_h\in \Gamma'} w_{y_h}\phi_{u_0}(\inner{y_h}{v}).
	\end{align*}
	It follows from the assumption that
	\begin{align*}
		\frac1{|V|}\sum_{v\in V}|G(v)-F_{w'}(v)|
		&\geq \frac{1}{|V|}\sum_{v\in V}|F_p(v)-F_{w'}(v)| - \frac{1}{|V|}\sum_{v\in V}|G(v)-F_p(v)|\\
		&\ge\kappa-\frac1{4\sqrt M}.
	\end{align*}
	Multiplying by $M$ proves the claimed inequality.
\end{proof}

Now we are ready to prove Lemma~\ref{lem:Y}.
\begin{proof}[Proof of Lemma~\ref{lem:Y}]
	Choose $c>0$ sufficiently small and then decrease $\bar\tau$ so that $N\geq 2$ and $q\ge 6$.
	Recall that $q$ is the smallest even integer such that $2^q\geq 512N$. By its minimality, $512N > 2^{q-2}$, so $512N \leq 2^q < 2048N$ and $256N \leq H = 2^{q-1} < 1024N$.
	
	Let the sets $Z$ and $V$ be as defined above.
	Lemma~\ref{lem:phi_u_kernel} gives $\sigma_{\min}(A) \geq c_2\sqrt H/q^{3/2}$.
	Applying Lemma~\ref{lem:probability_measure_Z}, we obtain a probability measure on $Z$ whose approximation error is at least
	\begin{equation}\label{eqn:evenmargin}
		\kappa
		= \frac{\sigma_{\min}(A)}{8H}
		\geq \frac{c_3}{\sqrt H\,q^{3/2}}
		\geq \frac{c_4}{\sqrt N\,q^{3/2}}.
	\end{equation}
	
	Let $\Lambda_\tau = \log(1/\tau)$. After decreasing $\bar\tau$ again, we obtain from the definition of $N$ that
	\[
	N\leq 2c\tau^{-2}\Lambda_\tau^{-3},\qquad q\lesssim \log N\lesssim \Lambda_\tau.
	\]
	Substituting these bounds into \eqref{eqn:evenmargin} yields
	\[
	\kappa\geq \frac{c_5}{\sqrt{c}\tau^{-1} \Lambda_\tau^{-3/2}\cdot \Lambda_\tau^{3/2}} = \frac{c_5}{\sqrt c}\tau
	\geq 4\tau,
	\]
	provided that $c$ is sufficiently small.
	For every integer $M\ge C\tau^{-2}$, with $C\ge 1$, Lemma~\ref{lem:unweighted_conversion} gives the required set	$\Gamma\subseteq\mathbb S^{q+2}$, since $\kappa-1/(4\sqrt M)\ge\tau$.
\end{proof}

%% file: vanilla.tex
\section{Lower Bound for Vanilla Clustering}\label{sec:vanilla}

The same construction gives a lower bound for vanilla $(k,z)$-clustering. For $k\leq\eps^{-(z+2)}$, it gives a cubic logarithmic loss, improving the quintic logarithmic loss in~\cite{HHLX26}. For $k > \eps^{-(z+2)}$, there is no logarithmic loss.

\begin{theorem}\label{thm:vanilla_LB}
	Let $z\geq 1$. There are constants $c,\eps_0 > 0$ depending only on $z$ such that, for every integer $k \geq 1$ and every $\eps\in (0,\eps_0]$, there exists a finite point set $X$ in a Euclidean space for which every weighted subset $S\subseteq X$ that is an $\eps$-coreset for $(k,z)$-clustering satisfies
	\[
	|S| \geq c\begin{cases}
		\displaystyle\frac{k^{2(z+1)/(z+2)}\eps^{-2}}{\log^3(1 + k^{-1/(z+2)}\eps^{-1})}, & k \leq \eps^{-(z+2)},\\
		k\eps^{-(z+2)}, & k > \eps^{-(z+2)}.
	\end{cases}
	\]
\end{theorem}

We shall use a vanilla version of Lemma~\ref{lem:construction_support}.
\begin{lemma}\label{lem:vanilla_support}
	Let $K,D,j$ be positive integers and let $R\geq 2z$ be real. For $0<\tau=A_0\eps R\leq\bar\tau$, let $X$ and $N$ be as in the construction of Lemma~\ref{lem:construction_support}. If $K+D\leq k$, $j\leq\min\{R^2,K/2\}$ and $K/2\leq jR^z\leq K$, then every $\eps$-coreset $S\subseteq X$ satisfies $|S|\geq KDN/4$.
\end{lemma}
\begin{proof}
	Suppose $|S|<KDN/4$. By the counting argument at the start of the proof of Lemma~\ref{lem:construction_support}, there is a set $J\subseteq[K]$ of size $j$ such that at least $jD/2$ cells $X_{i\ell}$ with $i\in J$ satisfy $|S\cap X_{i\ell}|\leq N$. Apply Lemma~\ref{lem:bad_directions} to these cells with $\calL=[D]$. Since $K+D\leq k$, Lemma~\ref{lem:two_center_sets} gives two sets $\calC_0,\calC_1$ of $k$ centers. Let $\delta(P)=\cost(P,\calC_0)-\cost(P,\calC_1)$ for $P\in\{X,S\}$. By the two lemmata,
	\[
		|\delta(X)-\delta(S)|\geq z u_0^{z-1}\frac{jDM\tau}{4}\geq \frac{A_0}{8}\eps n,
	\]
	where $n=KDM$ and the last inequality uses $u_0\geq R$, $\tau=A_0\eps R$ and $jR^z\geq K/2$.

	On the other hand, $\cost(X,\calC_1) \leq \cost(X,\calC_0) = (K-j)DM + u_0^z j D M \leq E_z n$, since $u_0\leq 3R/2$ and $jR^z\leq K$. The coreset guarantee thus gives $|\delta(X)-\delta(S)|\leq 2E_z\eps n$, contradicting $A_0>24E_z$.
\end{proof}

\begin{proof}[Proof of Theorem~\ref{thm:vanilla_LB}]
Take $\eps_0\leq\bar\tau/(A_0\lceil 2z\rceil)$ and let $k_0=4\lceil 2z\rceil^{z+2}$.
We first assume that $k \geq k_0$; smaller values of $k$ are handled at the end. Define
\[
R = \left\lfloor\min\left\{\left(\frac{k}{4}\right)^{\frac{1}{z+2}},\frac{\bar\tau}{A_0\eps}\right\}\right\rfloor,
\quad j = R^2,\quad K = \lceil R^{z+2}\rceil,\quad D = \left\lfloor \frac{k}{2}\right\rfloor,
\quad \tau = A_0\eps R,\quad M = \left\lceil \frac{C}{\tau^{2}}\right\rceil.
\]
Then $R \geq 2z$, $j \leq K/2 \leq jR^z \leq K \leq k/2$, $K+D \leq k$ and $\tau \leq \bar\tau$.
By Lemma~\ref{lem:vanilla_support}, every $\eps$-coreset $S$ of the instance $X$ satisfies $|S|\geq KDN/4$. Since $D=\lfloor k/2\rfloor\geq k/3$, the bound on $N$ in Lemma~\ref{lem:Y} gives
\[
|S| \geq \frac{1}{4}KDN \geq \frac{k}{12}KN
\gtrsim \frac{k R^{z+2}}{\tau^2\log^3(1/\tau)}
= \frac{k R^z\eps^{-2}}{A_0^2\log^3(1/\tau)}.
\]
Our choice of $R$ gives
\[
R \asymp_z \min\{k^{1/(z+2)},\eps^{-1}\},
\qquad \tau \asymp_z \min\{\eps k^{1/(z+2)},1\}.
\]
If $k \leq \eps^{-(z+2)}$, then $R \asymp_z k^{1/(z+2)}$ and
\[
\log(1/\tau) \asymp_z \log(1 + k^{-1/(z+2)}\eps^{-1}),
\]
which gives the first bound in the theorem. If $k > \eps^{-(z+2)}$, then $R \asymp_z \eps^{-1}$ and $\tau \asymp_z 1$. Thus $\log(1/\tau) = \Theta_z(1)$, and we obtain $|S| = \Omega_z(k\eps^{-(z+2)})$ without a logarithmic loss.

\paragraph{Small values of $k$.} It remains to consider $k < k_0$. Let $c_0 = c_0(z) > 0$ be a sufficiently small constant and decrease $\eps_0$ so that $\eps_0^2 \leq c_0/2$. Let $n = \lfloor c_0\eps^{-2}\rfloor \geq 2$ and $X = \{e_1,\dots,e_n\}\subseteq\R^{n+1}$.
Suppose that $S\subseteq X$ is a coreset with $|S| < n/2$ and let
\[
n_0 = n-|S|,\qquad v = \frac{1}{\sqrt{n_0}}\sum_{\substack{i\in[n]\\ e_i\notin S}}e_i.
\]
Both $\{v\}$ and $\{-v\}$ can be padded to exactly $k$ centers using distinct points $ae_{n+1}$ with $a \geq 3$, without changing the distance from any point of $X$ to its nearest center.
Every point of $S$ is orthogonal to $v$, so
\[
\cost(S,\{v\}) = \cost(S,\{-v\}).
\]
On the other hand,
\[
\cost(X,\{-v\}) - \cost(X,\{v\})
= n_0\left[\left(2 + \frac{2}{\sqrt{n_0}}\right)^{z/2} - \left(2 - \frac{2}{\sqrt{n_0}}\right)^{z/2}\right]
\gtrsim_z \sqrt{n_0} \geq \sqrt{n/2},
\]
where we used the mean-value theorem in the first inequality. 
The coreset guarantee and the equality of the two costs on $S$ give
\[
\cost(X,\{-v\}) - \cost(X,\{v\})
\leq \eps(\cost(X,\{-v\}) + \cost(X,\{v\}))
\leq 2^{z+1}\eps n.
\]
Combining these inequalities yields $1 \lesssim_z \eps\sqrt{n} \leq \sqrt{c_0}$, a contradiction for sufficiently small $c_0$. Hence $|S| \geq n/2 \gtrsim_z \eps^{-2}$. Since $k < k_0$, this proves the theorem after adjusting $c$.
\end{proof}